\documentclass[aps,pre,twocolumn,floats,showpacs]{revtex4}
\usepackage{epsfig}
\usepackage{color}
\usepackage{bm}
\usepackage{latexsym}

\begin{document}
\newcommand{\hide}[1]{}
\newcommand{\tbox}[1]{\mbox{\tiny #1}}
\newcommand{\half}{\mbox{\small $\frac{1}{2}$}}
\newcommand{\sinc}{\mbox{sinc}}
\newcommand{\const}{\mbox{const}}
\newcommand{\trc}{\mbox{trace}}
\newcommand{\intt}{\int\!\!\!\!\int }
\newcommand{\ointt}{\int\!\!\!\!\int\!\!\!\!\!\circ\ }
\newcommand{\eexp}{\mbox{e}^}
\newcommand{\bra}{\left\langle}
\newcommand{\ket}{\right\rangle}
\newcommand{\EPS} {\mbox{\LARGE $\epsilon$}}
\newcommand{\ar}{\mathsf r}
\newcommand{\im}{\mbox{Im}}
\newcommand{\re}{\mbox{Re}}
\newcommand{\bmsf}[1]{\bm{\mathsf{#1}}}
\newcommand{\mpg}[2][1.0\hsize]{\begin{minipage}[b]{#1}{#2}\end{minipage}}

\title{Distribution of $S$-matrix poles for one-dimensional disordered wires}
\author{I. F. Herrera-Gonz\'alez,$^1$ J. A. M\'endez-Berm\'udez,$^2$ and F. M. Izrailev$^{2,3}$}
\affiliation{$^1$Department of Engineering, Universidad Popular Aut\'onoma del Estado de Puebla, 21 Sur 1103, 
Barrio Santiago, Puebla, Pue., Mexico}
\affiliation{$^2$Instituto de F\'{\i}sica, Benem\'erita Universidad Aut\'onoma de Puebla,
Apartado Postal J-48, Puebla 72570, Mexico}
\affiliation{$^3$NSCL and Dept.~of Physics and
Astronomy, Michigan State University - East Lansing, Michigan
48824-1321, USA}

\date{\today}

\begin{abstract}

By the use of the effective non-Hermitian Hamiltonian approach to scattering we study the distribution 
of the scattering matrix ($S$-matrix) poles in one-dimensional (1D) models with various types of diagonal 
disorder. We consider the 
case of 1D tight-binding wires, with both on-site uncorrelated and correlated disorder, coupled to the
continuum through leads attached to the wire edges. In particular, we focus on the location of the 
$S$-matrix poles in the complex plane as a function of the coupling strength and the disorder strength. 
Specific interest is paid to the super-radiance transition emerging at the perfect coupling between wire 
and leads. We also study the effects of correlations intentionally imposed to the wire disorder. 
\end{abstract}


\maketitle

\section{Introduction}

To date, the paradigmatic one-dimensional (1D) Anderson model with white-noise diagonal disorder has 
been rigorously studied in great detail. The main result is that all eigenstates are exponentially localized 
in the infinite geometry, characterized by the amplitude decrease with the distance from their centers. 
The characteristic scale on which they are effectively localized is known as the localization length 
$L_{\mbox{\scriptsize{loc}}}$ 
which can be easily computed numerically in the frame of the transfer matrix method (see Ref.~\cite{IKM12} 
and references therein). This energy-dependent length is of utmost importance due to the single parameter 
scaling, according to which all transport properties of the finite wires are explicitly defined by the ratio 
$L_{\mbox{\scriptsize{loc}}}/N$, where $N$ is the length of the wire (see, for instance, Ref.~\cite{LGP88}). 
Specifically, 
there are various rigorous approaches allowing to derive the distribution function for the transmission 
coefficient characterizing the scattering process through finite wires of size $N$ (for references, 
see~\cite{M99}).   

Complimentary to the transfer matrix method, the scattering properties of electromagnetic waves 
propagating through finite wires can also be studied via the scattering matrix ($S$-matrix). There 
is an enormous number of papers devoted to the theory of the $S$-matrix in application to complex 
physical systems such as heavy nuclei, many-electron atoms, and quantum dots. Assuming a quite 
complex (chaotic) behavior of the closed (isolated) systems, it was suggested that various types of 
random matrices can serve as good mathematical models in describing statistical properties of 
scattering. In particular, typical models are represented 
by non-Hermitian matrices with a real part in terms of a fully random matrix plus an imaginary part of 
certain structure absorbing the details of the coupling to the continuum. In this way many results both 
analytical and numerical, have been obtained during the last decades (see, for example, 
Refs.~\cite{MW69,supersym,SZ92} and references therein). 

One important question in scattering theory is about the type of distribution of the widths of resonances, 
emerging in the transmission coefficient as a function of the energy. For sufficiently week coupling to the 
continuum (slightly open systems) the resonances are well isolated from each other; a situation termed as
{\it isolated resonances}. This happens when the resonance widths are smaller than the spacings between 
the locations of resonances. A completely different situation arises when the widths of the resonances 
are much larger than the spacing between them. In this case the resonances are strongly overlapped, 
thus resulting in specific properties of scattering. For random matrix models it was shown that the 
crossover from isolated to overlapped resonances is quite sharp in dependence on the coupling strength 
$\gamma$, and at the transition point the mean value of the resonances diverges. With the increase of 
$\gamma$ towards strongly overlapped resonances, a remarkable effect emerges, known as 
{\it superradiance} \cite{SZ92}. In this regime a finite number of resonances have very large widths while
the other resonances begin to be more and more narrow with the increase of the coupling.

The goal of this paper is to study the 1D Anderson model with both white-noise disorder and correlated 
disorder, focusing on the pole distribution in dependence of the degree of localization of eigenstates and 
of the strength of the coupling to the continuum. Although this model is very different from random 
matrix models, we expect that some of the properties, such as the superradiant transition and the 
divergence of the resonance widths at the superradiant transition point, develop similarly to those in 
random matrix theory (RMT) models. The distinctive property of our model is that by increasing the 
disorder one can change the degree of localization of eigenstates in the closed wires, therefore, in 
the wires attached to the leads the localization effects may greatly affect the pole distribution. Another 
issue to address is how correlations imposed to the diagonal disorder modify the pole distribution at the 
mobility edges emerging due to specific long-range correlations. We hope that our numerical 
results can help to develop an analytical approach to the problem of scattering in 1D wires in the case 
when a closed system is strongly influenced by localization effects.

\section{The Model and Scattering setup}

The 1D Anderson model is defined by the stationary Schr\"odinger equation 
\begin{equation}
\psi_{n+1}+\psi_{n-1}+\epsilon_n\psi_n=E\psi_n \ .
\label{Eq1} 
\end{equation}
for the electron wave function $\psi_n$ of energy $E$. In what follows we consider both uncorrelated and 
correlated disorder specified by the site potentials $\epsilon_n$. The energies $E$ and $\epsilon_n$ are 
dimensionless quantities measured in units of the kinetic electron energy. 

In the non-disordered case ($\epsilon_n=0$), the solutions $\psi_n$ are plane waves with wave number 
$\mu$ defining the dispersion relation
\begin{eqnarray}
E=2\cos \mu \ , \quad 0\le \mu \le \pi \ .
\label{Eq2}
\end{eqnarray}
For the the disordered case we assume that the distribution of on-site energies $\epsilon_n$ is characterized 
by random variables with zero mean and variance $\sigma^2$,
\begin{eqnarray*}
\langle \epsilon_n \rangle=0 \ , \quad \langle \epsilon^2_n \rangle=\sigma^2 \ .
\end{eqnarray*}
Here $\langle \dots \rangle$ stands for the average over disorder realizations. We also assume that the 
disorder is weak, 
\begin{equation}
\sigma^2\ll 1 \ ,
\label{Eq3}
\end{equation}
which is needed to develop a proper perturbation theory.

In the case of uncorrelated disorder all electron eigenstates are exponentially localized with the 
characteristic length $L_{\mbox{\scriptsize{loc}}}$ in the limit of infinite wire length  ($N\rightarrow \infty$). 
As is known, for white-noise disorder the localization length $L_{\mbox{\scriptsize{loc}}}$ is given 
by the Thouless expression \cite{T79} (see also Ref.~\cite{IKM12}),
\begin{equation}
L^{-1}_{\mbox{\scriptsize{loc}}}=\frac{\sigma^2}{8\sin^2\mu}=\frac{w^2}{96(1-E^2/4)} \ .
\label{loc}
\end{equation}
Here $\mu$ is defined through the dispersion relation (\ref{Eq2}) and the second relation is given 
for a box distribution of $\epsilon_n$ specified by the interval $[-w/2,w/2]$.  

Below we study 1D finite wires of size $N$ with disordered on-site potentials $\epsilon_n$. 
The first and last sites of the disordered lattice are connected to semi-infinite {\it ideal} leads 
through coupling amplitudes $\sqrt{\gamma}$. In this way the leads are considered as a 
continuum to which the disordered wire is coupled according to given boundary conditions. 
The scattering properties of such open system can be formulated in terms of the non-Hermitian 
Hamiltonian \cite{MW69,SZ92}. The key point of this approach is based on the projection of the 
total Hermitian Hamiltonian (disordered wire plus leads) onto the basis defined by 
the Hamiltonian $H$ describing the properties of the closed model (the disordered 
wire only). 
 
In our model, the non-Hermitian Hamiltonian has the following form near the band center ($E=0$) 
\cite{SIZC12}:
\begin{equation}
{\cal H}_{mn}=H_{mn}-\frac{i}{2}W_{mn}
\label{Eq4}
\end{equation}    
with
\begin{equation}
W_{mn}=2\pi \sum_{c=L,R} A^{(c)}_mA^{(c)}_n  \ ,
\label{Eq5}
\end{equation}
where $W_{mn}$ is defined by the coupling amplitudes
\begin{equation}
A^{L,R}_i=\sqrt{\frac{\gamma}{\pi}}\left(\delta^{(L)}_{i1}+\delta^{(R)}_{iN}\right) \ .
\label{Eq6}
\end{equation}
Here, $H_{mn}$ is the Hamiltonian of the 1D Anderson model
\begin{equation}
H_{nm}=\epsilon_n \delta_{mn}+\delta_{m,n+1}+\delta_{m,n-1} \ ,
\label{Eq7}
\end{equation}
while the non-Hermitian part $W$ is given in terms of the coupling amplitudes $A^{(c)}_i$ 
between the internal states $|i\rangle$ and open decay channels $c=L,R$, where $L$ and 
$R$ stand for left and right, respectively.

With the non-Hermitian Hamiltonian, it is possible to obtain the scattering matrix $S$ in the form,
\begin{equation}
S=\frac{1-iK}{1+iK} \ ,
\label{Eq8} 
\end{equation}
where the reaction matrix $K$ is given by 
\begin{eqnarray}
K^{ab}=\sum_j \frac{C^{(a)}_jC^{(b)}_j}{E-E_j}, \ \  C^{(c)}_j=\sum_mA^{(c)}_m\psi^{(j)}_m \ ,
\label{Eq9}
\end{eqnarray}
and $\psi^{(j)}_m$ is the $m$ component of the $j$-th eigenstate of the closed Hamiltonian 
(\ref{Eq7}) with eigenvalues $E_j$.

One of the quantities studied in this work is the distribution of the eigenvalues 
\begin{equation}
\Omega_k=\omega_k-\frac{i}{2}\Gamma_k
\label{poles}
\end{equation}
of the non-Hermitian Hamiltonian (\ref{Eq4}); here 
$\omega_k$ and $\Gamma_k$ are called the position and the width of the $k-$resonance, 
respectively. The eigenvalues $\Omega_k$ can also be treated as the poles of the $S$-matrix, 
since one can write
\begin{eqnarray*} 
S(E)=1-2\pi iA^{T}(E)\frac{1}{E-{\cal H}}A(E) \ ,
\end{eqnarray*}
where $A(E)$ is the $N\times 2$ matrix with columns composed by the coupling amplitudes $A^{(c)}_i$.
   
In general the poles of the $S$-matrix for the 1D Anderson model depend 
on three parameters: the energy $E$, the coupling constant $\gamma$, and the localization length 
$L_{\mbox{\scriptsize{loc}}}$ (given by Eq.~(\ref{loc})). However, without loss of generality, in this work 
we fix the energy to $E=0$. It is known that depending on the 
ratio between the localization length and the system size $N$, there are three different regimes for the 
scattering processes: the ballistic regime characterized by $L_{\mbox{\scriptsize{loc}}}\gg N$, the chaotic 
regime where $L_{\mbox{\scriptsize{loc}}}\approx N$, and the localized regime which occurs for 
$L_{\mbox{\scriptsize{loc}}}\ll N$ (see for example Ref.~\cite{IKM12}). 

%%%%%%%%%%%%%%%%%%%%%%%%%%%%%%%%%%%%%%%%%%%%%%%%%%%%%%%%%%%%%%%%%%%%%%%%%%%%%%%%%%%%
\begin{figure*}[!htp]
\includegraphics[width=5.5cm,height=4cm]{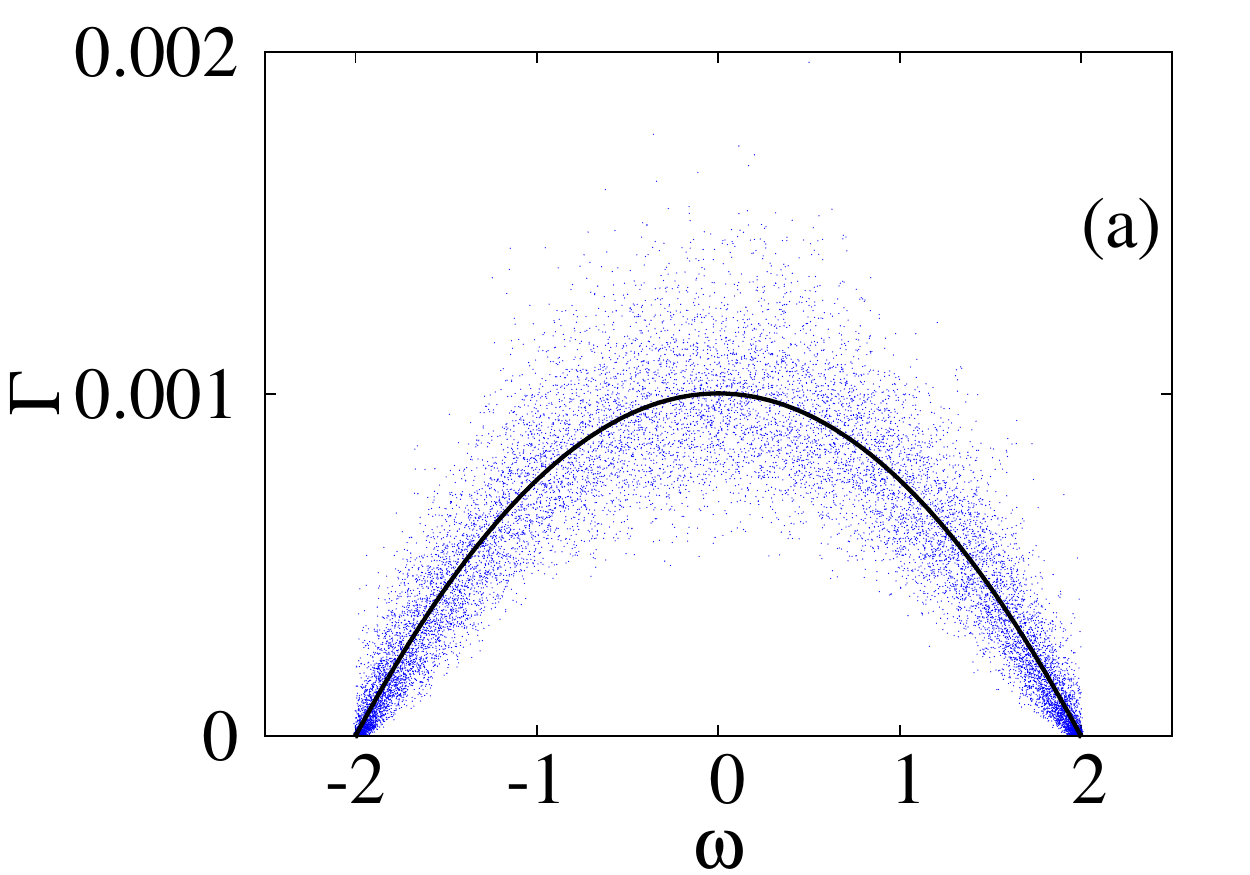}
\includegraphics[width=5.5cm,height=4cm]{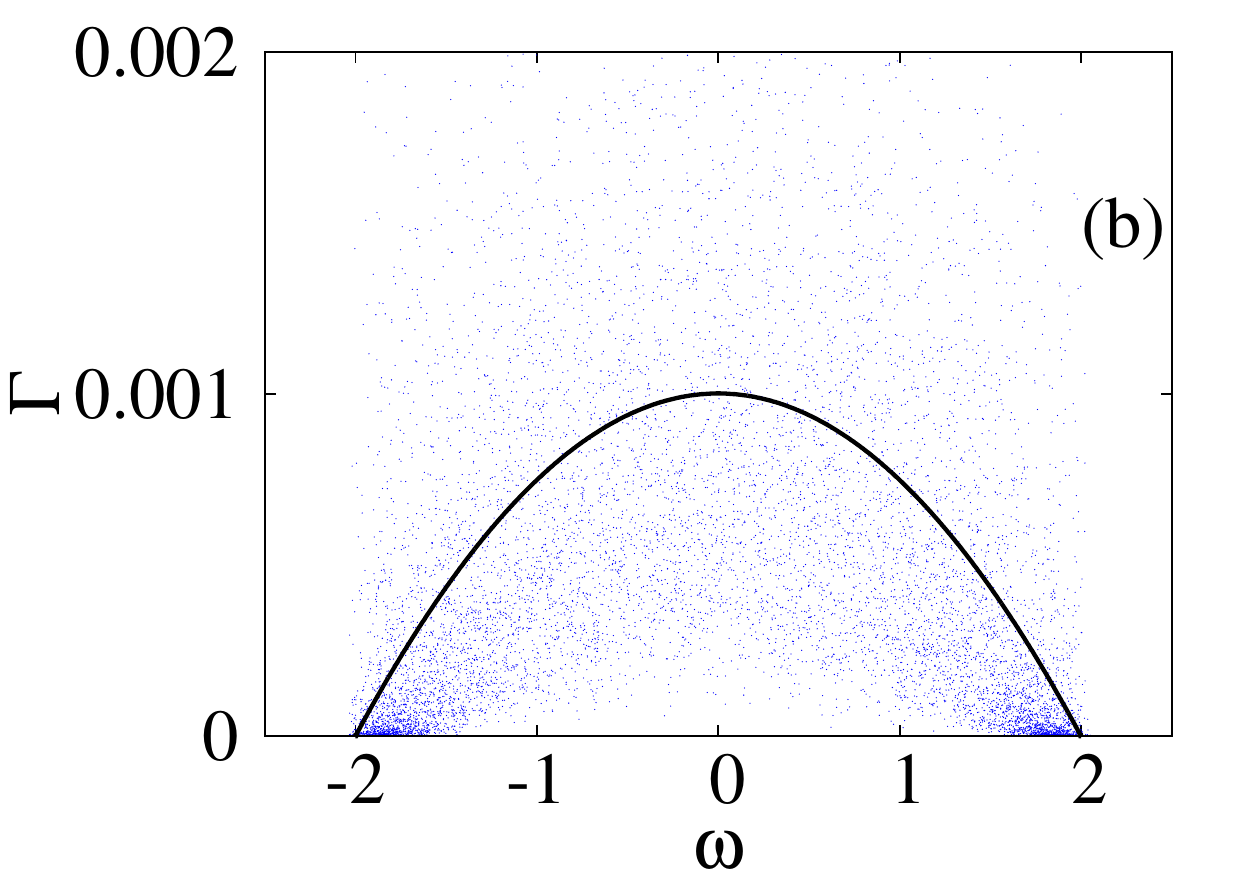}
\includegraphics[width=5.5cm,height=4cm]{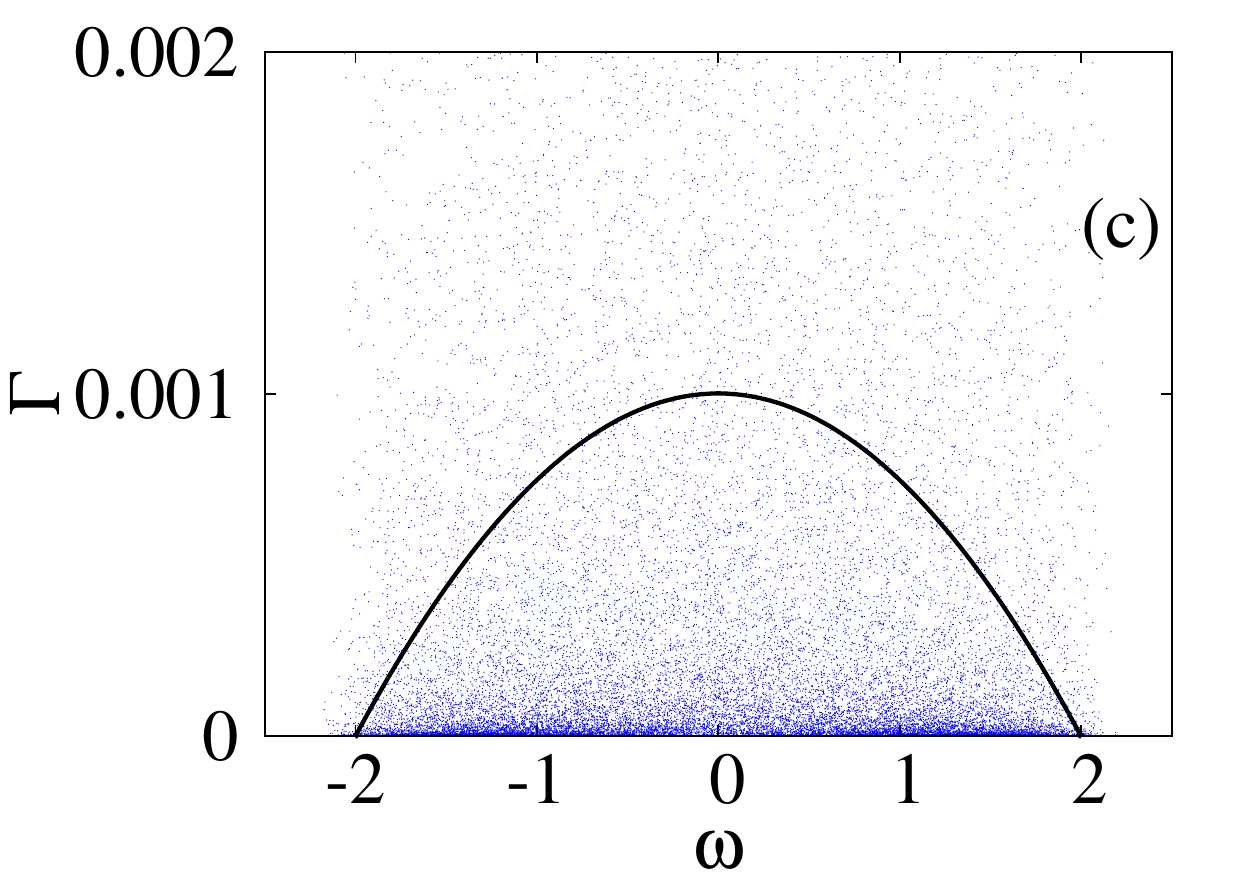}
\includegraphics[width=5.5cm,height=4cm]{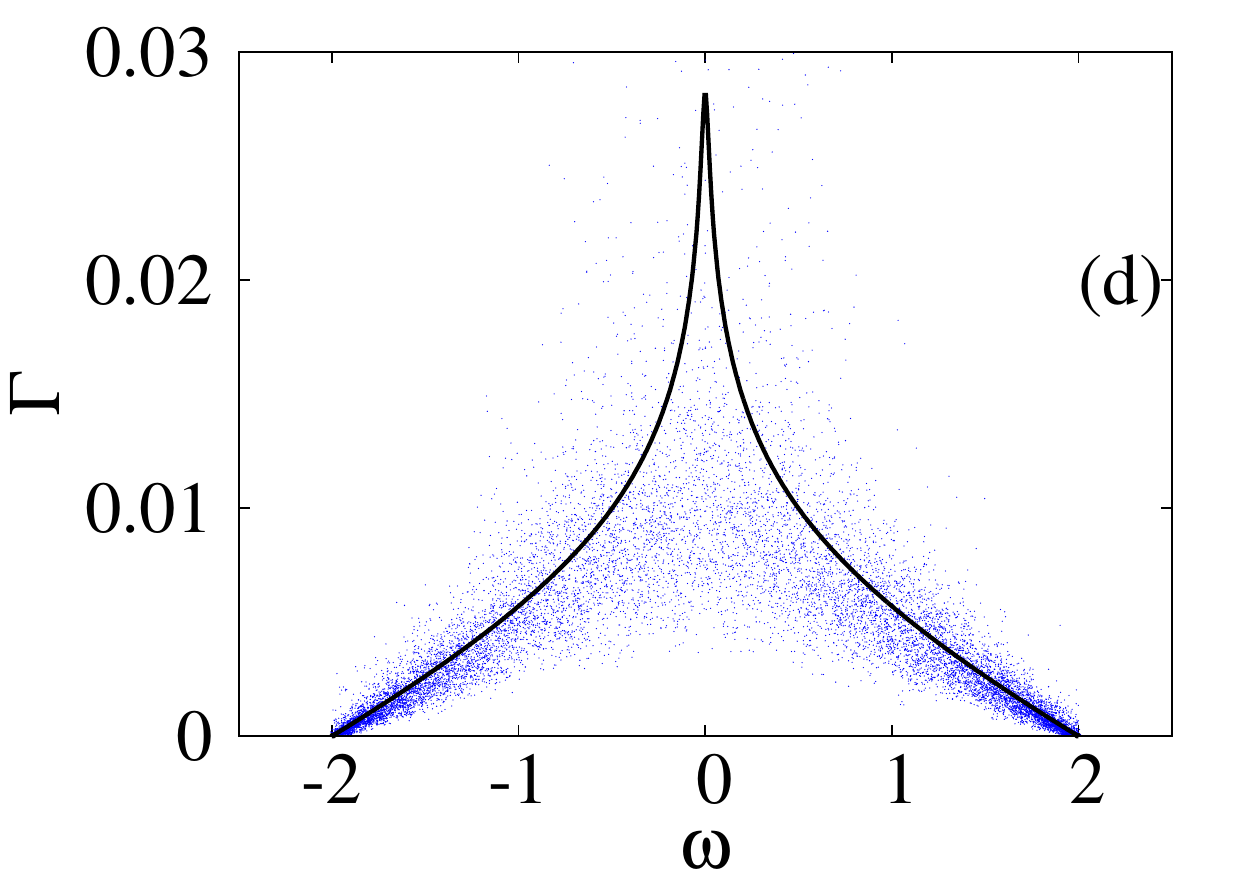}
\includegraphics[width=5.5cm,height=4cm]{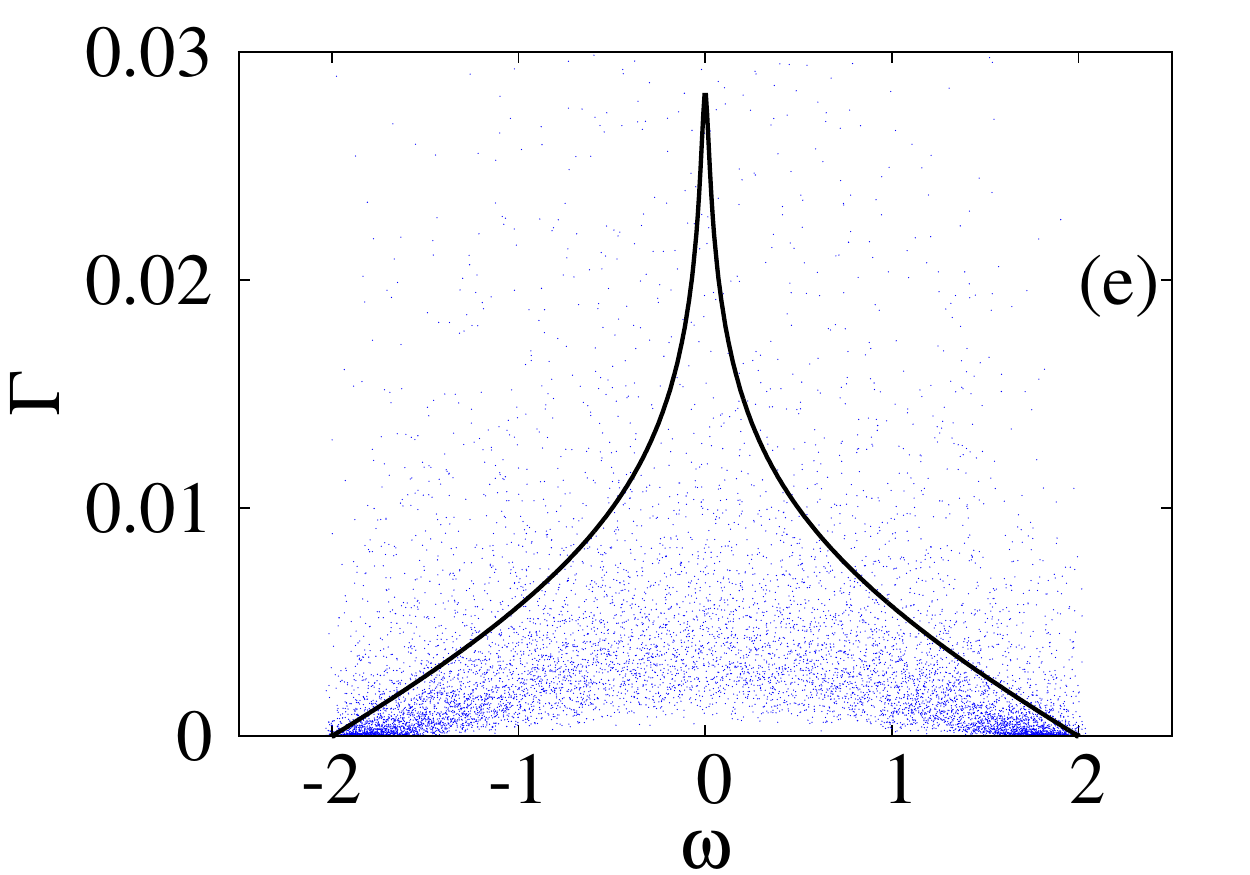}
\includegraphics[width=5.5cm,height=4cm]{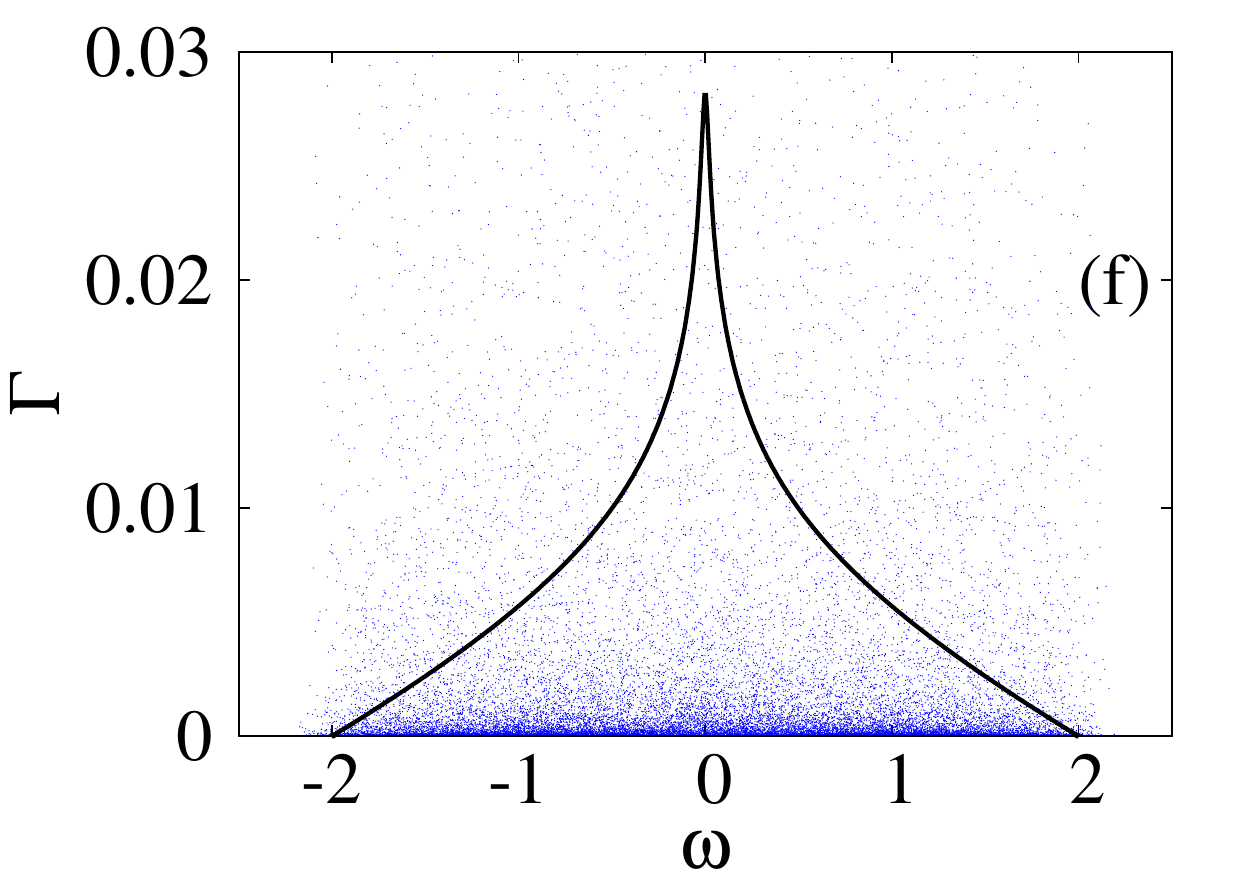}
\caption{\label{Fig1} Imaginary vs.~real part of the $S$-matrix poles $\Omega$ [see Eq.~(\ref{poles})] 
for 1D disordered wires of length $N=800$ coupled to the continuum with strength $\gamma=0.1$ 
(upper panels) and $\gamma=1$ (lower panels). The disorder strength was set to 
$\sigma^2=0.001$, $L_{\mbox{\scriptsize{loc}}}/N=10$ (left panels); 
$\sigma^2=0.01$, $L_{\mbox{\scriptsize{loc}}}/N=1$ (middle panels); and 
$\sigma^2=0.1$, $L_{\mbox{\scriptsize{loc}}}/N=0.1$ (right panels). 
Black curves represent the corresponding non-disordered wires. Here, 50 wire realizations were used.} 
\end{figure*}
%%%%%%%%%%%%%%%%%%%%%%%%%%%%%%%%%%%%%%%%%%%%%%%%%%%%%%%%%%%%%%%%%%%%%%%%%%%%%%%%%

%%%%%%%%%%%%%%%%%%%%%%%%%%%%%%%%%%%%%%%%%%%%%%%%%%%%%%%%%%%%%%%%%%%%%%%%%%%%%%%%%
\begin{figure*}[!htp]
\includegraphics[width=5.5cm,height=4.5cm]{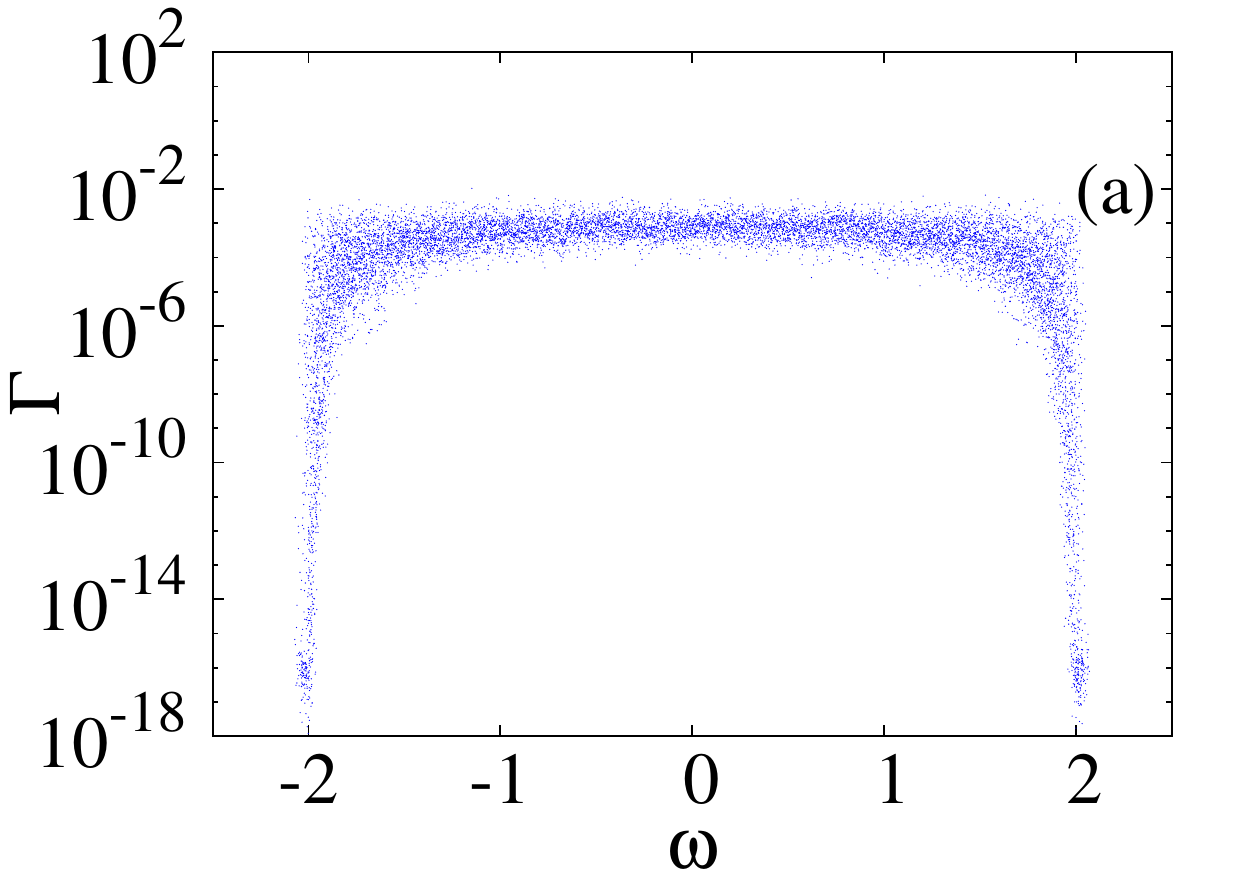}
\includegraphics[width=5.5cm,height=4.5cm]{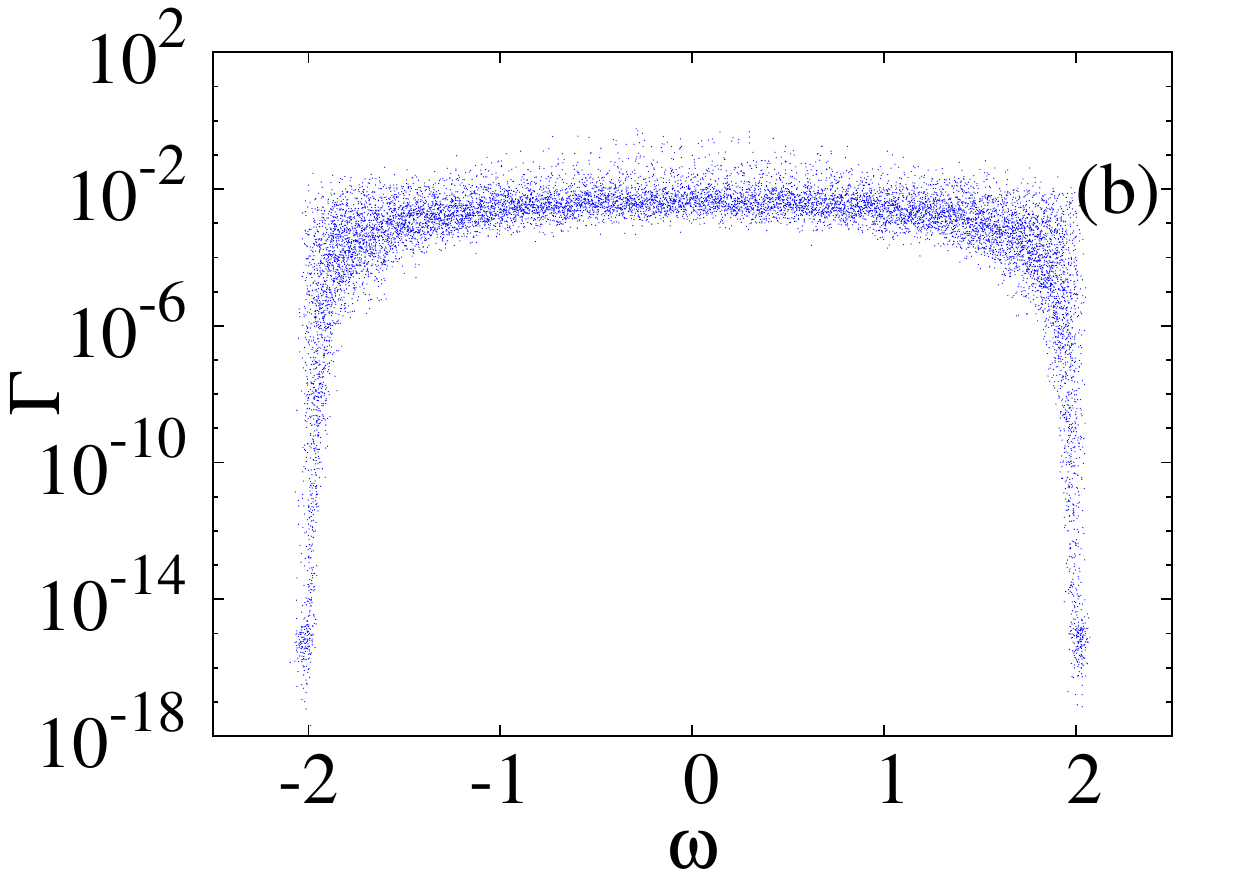}
\includegraphics[width=5.5cm,height=4.5cm]{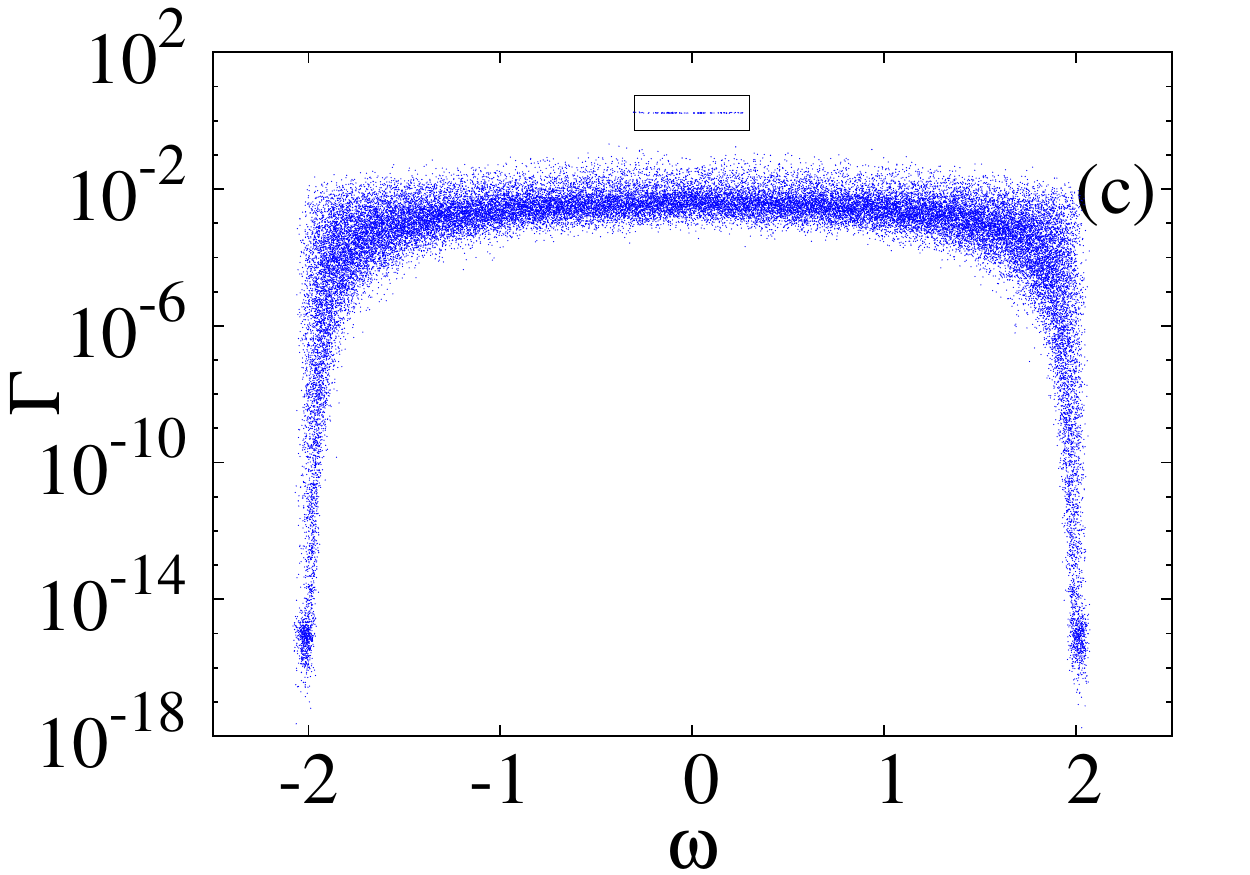}
\caption{\label{Fig2} Imaginary vs.~real part of the $S$-matrix poles $\Omega$ [see Eq.~(\ref{poles})] 
for 1D disordered wires of length $N=800$ coupled to the continuum with (a) $\gamma=0.1$, 
(b) $\gamma=1$, and (c) $\gamma=1.5$. The disorder strength was set to $\sigma^2=0.01$ such that 
$L_{\mbox{\scriptsize{loc}}}/N= 1$. Here, 50 wire realizations were used. The small rectangle in (c) 
encloses the superradiant states.} 
\end{figure*}
%%%%%%%%%%%%%%%%%%%%%%%%%%%%%%%%%%%%%%%%%%%%%%%%%%%%%%%%%%%%%%%%%%%%%%%%%

\section{$S$-matrix poles: White-noise disorder}

In this Section we consider the case of uncorrelated disorder: 
$\langle \epsilon_n \epsilon_{n+m} \rangle=0$, for $m\ne 0$. First, in Fig.~\ref{Fig1} we report the $S$-matrix poles 
as the localization length decreases (from left to right) for two different values of the coupling strength: 
$\gamma=0.1$ (upper panels) and $\gamma=1$ (lower panels). In the ballistic regime, 
$L_{\mbox{\scriptsize{loc}}}\gg N$, see Fig.~\ref{Fig1}(left panels), the poles are distributed around 
the curves corresponding to the non-disordered case, which are shown as continuous black curves in 
all panels of Fig.~\ref{Fig1}. This behavior is easy to understand since the eigenstates of the disordered
wires in this regime remain close to the eigenstates of wires with zero disorder (plane waves). In the 
chaotic regime, $L_{\mbox{\scriptsize{loc}}}\approx N$, the eigenstates are still extended (as in the 
ballistic regime), however, they produce strong fluctuations of $\Gamma$ values. As one can see by 
comparing Fig.~\ref{Fig1}(left and middle panels), the distribution of poles is 
quite sensitive to whether the eigenstates are quasi-regular or chaotic. Finally, for the localized regime, 
$L_{\mbox{\scriptsize{loc}}}\ll N$, only few eigenstates touch the wire boundaries and, 
as a consequence, the influence of the continuum is reduced producing a distribution of 
poles closer to the real axis, as Fig.~\ref{Fig1}(right panels) shows.

As the coupling parameter $\gamma$ increases, we observe the following effect. At zero coupling 
to the continuum the $S$-matrix poles are located along the real axis. As the coupling is turned on 
the poles acquire an imaginary part and, if the average width $\langle \Gamma \rangle$ is small as 
compared to the level spacing $D$ of the closed system, the cross sections in the scattering process 
reveal isolated resonances and the poles form a single cloud close to the real axis in the complex plane. 
However, with the increase of the coupling parameter $\gamma$, 
a crossover from isolated to overlapping resonances occurs; this crossover at $\gamma\approx 1$ 
is characterized by the appearance of two clouds of poles in the complex plane: one cloud
corresponds to isolated resonances (with small $\Gamma$; $\Gamma \ll D$) and the other one to 
strongly overlapped ones (with large $\Gamma$; $\Gamma \gg D$). The latter states are termed 
superradiant states since they are short-lived, in contrast with long-lived states with small $\Gamma$. 
In the literature this segregation of poles is known as the {\it superradiant} transition. As we 
demonstrate below, the transition between isolated and superradiant
states is very sharp with respect to the change of $\gamma$ and can be associated with a kind of 
phase transition.  

Then, in Fig.~\ref{Fig2} the aforementioned pole segregation is displayed. 
The lower cloud in Fig.~\ref{Fig2}(c) corresponds to $(N-2)N_d$ long-lived states, 
whereas the upper cloud (labeled by the small rectangle) represents $2N_d$ 
short-lived or superradiant states, being $N_d$ 
the number of realizations of the disorder (the factor 2 here accounts for the number of leads
connected to the 1D wire). Notice that in this figure we show the superradiant transition for the 
chaotic regime $L_{\mbox{\scriptsize{loc}}}\approx N$ only, however, the existence of such 
transition and the value of $\gamma$ where it takes place do not depend on the degree of 
localization.

It is important to note that the control parameter determining the strength of the coupling to the 
continuum can be written as \cite{SIZC12}
\begin{eqnarray*}
\kappa =\frac{2\pi \gamma}{ND} \ ,
\end{eqnarray*}
where $D$ is the mean level spacing at the center of the energy band of the isolated wire. 
Note that $D$ can easily be evaluated if one takes into account the weak disorder condition 
(\ref{Eq3}). In this situation, the eigenvalues of the Hamiltonian (\ref{Eq7}) of the 
disordered wire are practically equal to the corresponding eigenvalues of the Hamiltonian
(\ref{Eq7}) with $\epsilon_n=0$. Therefore, the dispersion relation (\ref{Eq2}) can be used. In 
addition, if fixed boundary conditions are imposed, the wave number takes the discrete values 
$\mu_q=q\pi/(N+1)$,  with $q=1,\dots N$, and $D$ is then simply given by
\begin{equation}
D=\frac{2\pi}{N} \ .
\label{Eq11}
\end{equation}
Therefore, near to the band center $\kappa \approx \gamma$ and the superradiant transition 
takes place at $\kappa \approx 1$. It is quite interesting that the same critical value of 
$\kappa$ emerges also in other models such as the Gaussian Orthogonal Ensemble (GOE) 
of random matrices and two-body random interaction models \cite{brody81,TBRI}. The 
superradiant transition in the 1D Anderson model has already been reported in the literature, 
see for example Refs.~\cite{VZ05,celardo10}. In addition, the interplay between supperradiance 
and disorder has been previously established when all the sites of a lattice are coupled to a 
common decay channel \cite{CBKB13}. Also, the pole distribution, at perfect coupling, for the 
tree-dimensional Anderson model has been studied in \cite{WMK06}.

\section{Mean value and fluctuations of resonance widths} 

In the previous Section, we made a qualitative description of the superradiant transition by analyzing 
the distribution of poles of the $S$-matrix in the complex plane. Now we focus on how the mean value 
of resonances (more precisely, the mean value of the imaginary part of the eigenvalues) depends on 
the strength of the coupling to continuum. This problem has been studied in detail for non-Hermitian 
Hamiltonians, see Eq.~(\ref{Eq4}), in which the real part $H$ is a full random matrix belonging to one 
of standard ensembles (for example, to the GOE), and the imaginary part $W$ describes the coupling 
to continuum through a finite number of channels according to Eq.~(\ref{Eq5}); see for example 
\cite{CIZB08} and references therein.

%%%%%%%%%%%%%%%%%%%%%%%%%%%%%%%%%%%%%%%%%%%%%%%%%%%%%%%%%%%%%%%%%%%%%%%%%%%
\begin{figure*}[!htp]
\includegraphics[width=5.5cm,height=4.5cm]{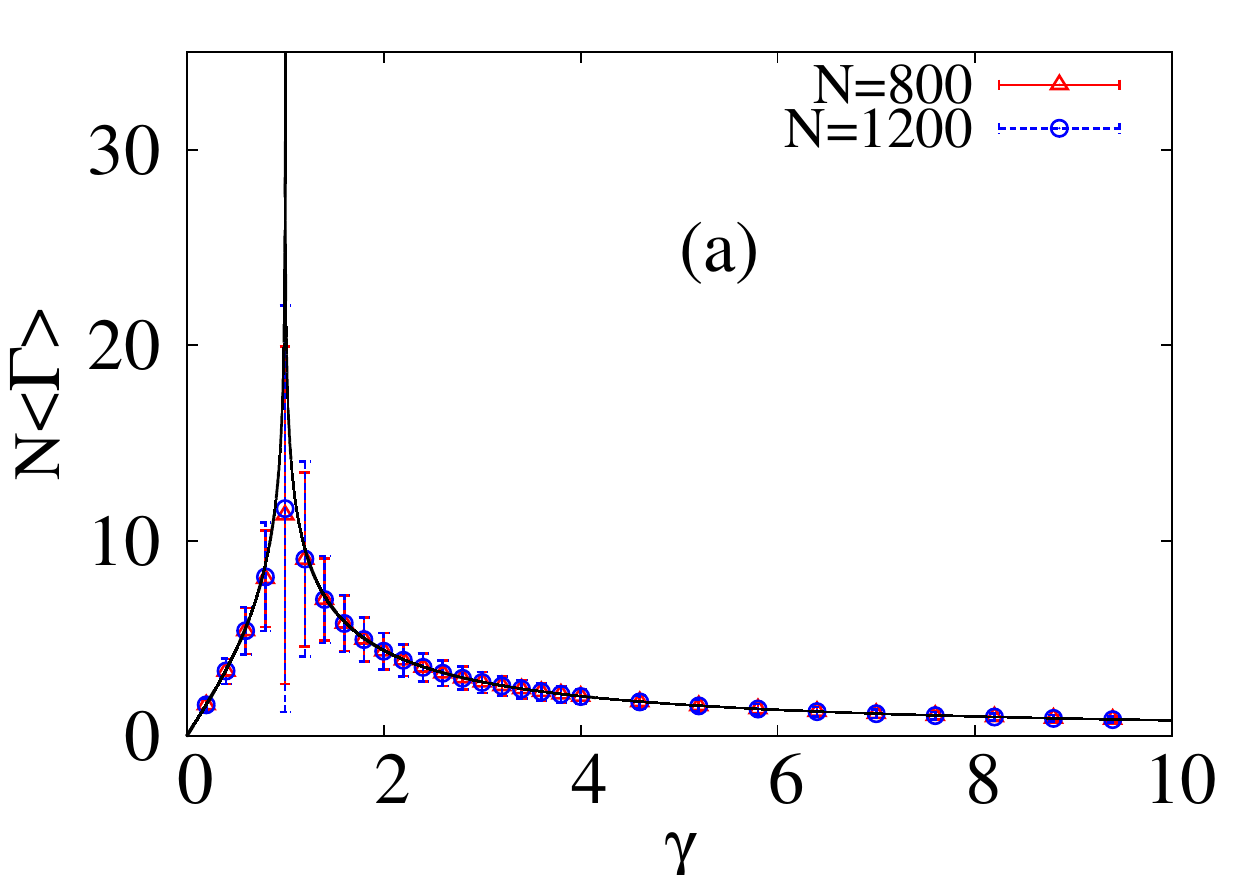}
\includegraphics[width=5.5cm,height=4.5cm]{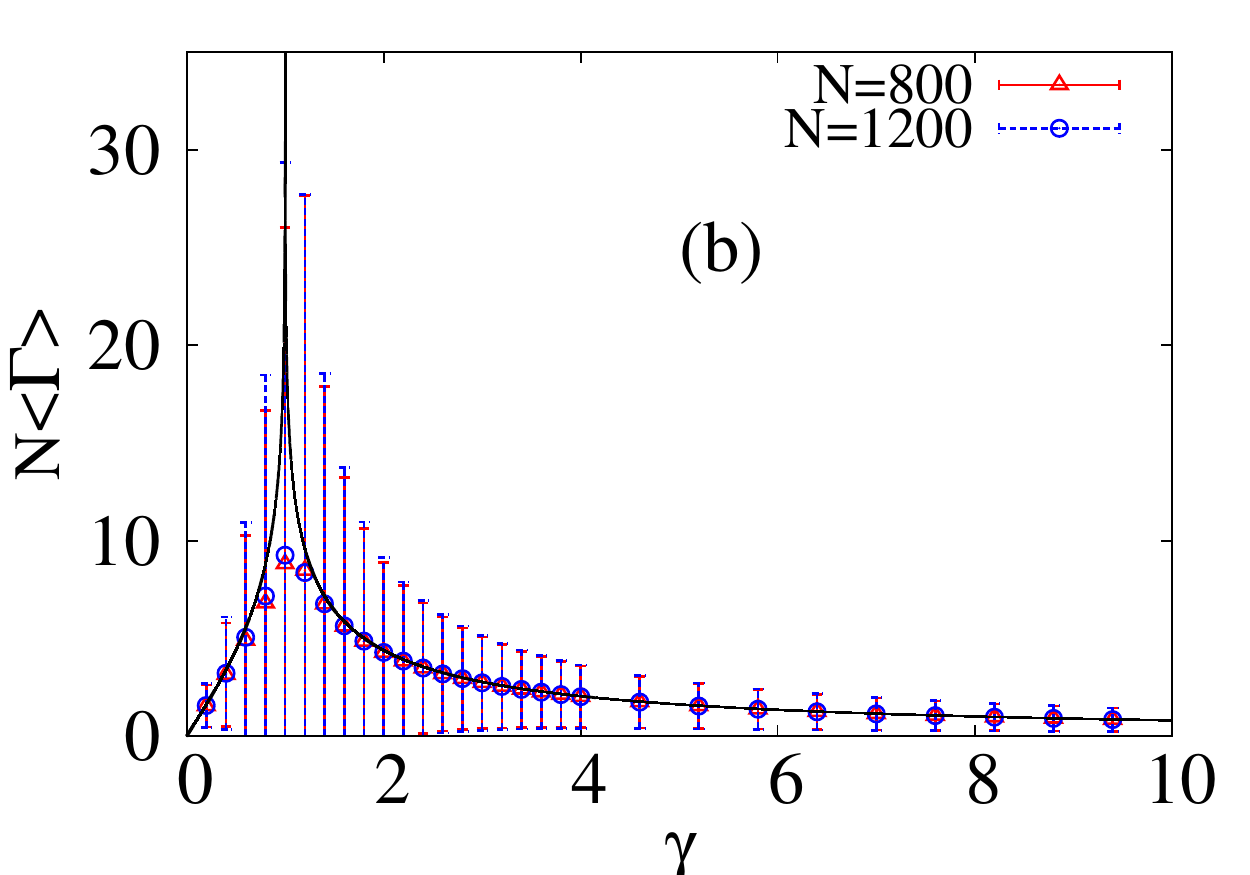}
\includegraphics[width=5.5cm,height=4.5cm]{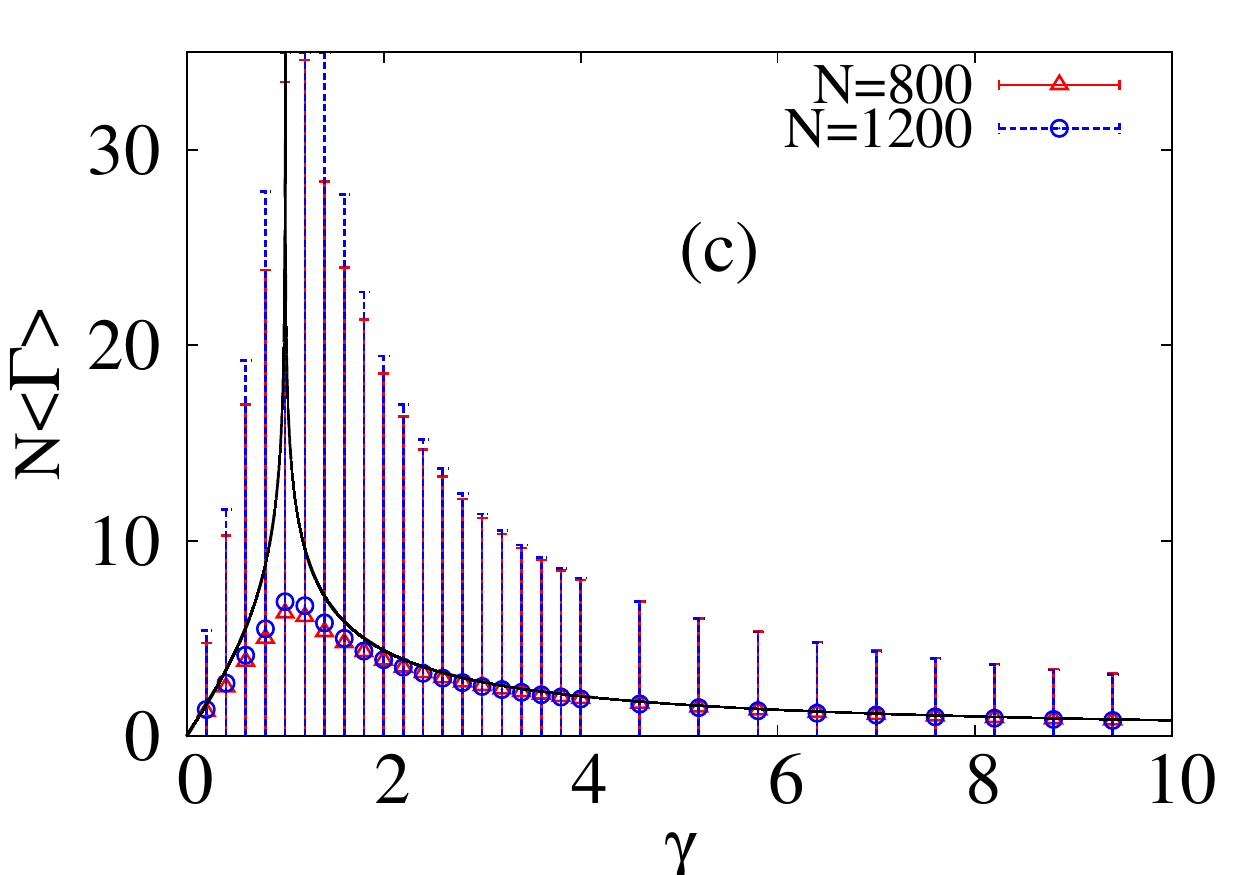}
\caption{\label{Fig3}  Average resonance width $\langle\Gamma\rangle$ at the band center as a 
function of the coupling to the continuum $\gamma$. Error bars are the corresponding standard 
deviation. (a) $L_{\mbox{\scriptsize{loc}}}/N=10$ (ballistic regime), (b) $L_{\mbox{\scriptsize{loc}}}/N=1$ 
(chaotic regime), and (c) $L_{\mbox{\scriptsize{loc}}}/N=0.1$ (localized regime). Circles and triangles 
represent different wire sizes. Black curves are Eq.~(\ref{Eq12}). Here, 50 wire realizations were used.}
\end{figure*}
%%%%%%%%%%%%%%%%%%%%%%%%%%%%%%%%%%%%%%%%%%%%%%%%%%%%%%%%%%%%%%%%%%%%%%%%%%%%%%%%

One important analytical result, in the case of $M$ channels $c=1,..., M$, is that the mean width of 
the resonances reads \cite{FS96} 
\begin{eqnarray}
\langle\Gamma\rangle=-\frac{MD}{2\pi}\ln \left(\frac{\tau-1}{\tau+1}\right), \ \ 
\tau=\frac{1}{2}\left(\gamma+\gamma^{-1}\right).
\label{Eq10}
\end{eqnarray} 
Here $D$ stands for the mean energy level spacing of the closed system at the band center $E=0$. 
Relation (\ref{Eq10}) is known as the Moldauer-Simonius equation which is widely used in 
physics \cite{MS67}. In the case when $H$ is a member of the Gaussian Unitary Ensemble 
(GUE) of random matrices (with $M$ equivalent $c-$channels) the analytical result for the whole 
distribution of individual widths $\Gamma_i$ has been derived in Ref.~\cite{F17}. The logarithmic 
divergence of $\langle\Gamma\rangle$ at the critical coupling $\tau=1$ is a direct consequence of the 
power law decay of large values of $\Gamma_i$. However this rigorous result refers to an infinite 
number of resonances, and for finite $N$ one has to take into account that $\langle\Gamma\rangle$ 
remains finite for any value of $\tau$, including $\tau=1$. 

In the case of the 1D Anderson model, where the Hamiltonian $H$ is given by the tridiagonal matrix 
of Eq.~(\ref{Eq7}) with diagonal disorder, a rigorous expression for $\langle\Gamma\rangle$ is unknown. 
However, our expectation is that relation (\ref{Eq10}) may also be applied to the 1D Anderson model. 
The physical argument for this expectation is that it may not be relevant whether a closed system, 
described by $H$, is a one-body or a many-body system. We expect whether this argument is 
valid only when the eigenstates of $H$, describing the closed system, are fully chaotic. As we show, 
strong differences occur for the model with strong disorder leading to localized eigenstates. 

With the use of expression (\ref{Eq11}) for $D$, we arrive to the following relation for the mean width 
of resonances: 
\begin{equation}
N\langle\Gamma\rangle=-2\ln \left(\frac{\tau-1}{\tau+1}\right) \, ,
\label{Eq12}
\end{equation} 
where $M=2$ is explicitly used. Notice that Eq.~(\ref{Eq12}) is invariant with respect to the 
change $\gamma \rightarrow 1/\gamma$. In fact, one can show that the whole distribution of poles 
corresponding to long-lived states is invariant under such a change. Moreover, transport properties 
in the 1D Anderson model have been shown to be still symmetric under the above change 
\cite{SIZC12}. This symmetry has been observed if $H_{mn}$ in Eq.~(\ref{Eq7}) is replaced by full 
random matrices (see e.g.~Refs.~\cite{SZ92,FS96}).
   
The validity of Eq.~(\ref{Eq12}) is confirmed in Fig.~\ref{Fig3} for the three regimes (ballistic, chaotic, 
and localized). We observe an excellent agreement between Eq.~(\ref{Eq12}) and the numerical data 
except for the points in the vicinity of $\gamma=1$, where differences are due to finite size effects. 
Surprisingly, the mean average width is practically insensitive to the degree of disorder if $\gamma$ 
is not too close to $\gamma=1$. 

%%%%%%%%%%%%%%%%%%%%%%%%%%%%%%%%%%%%%%%%%%%%%%%%%%%%%%%%%%%%%%%%%%%%%%%%%%%%%%%%%%%%%
\begin{figure}[!htp]
\begin{center}
\includegraphics[width=7cm,height=5cm]{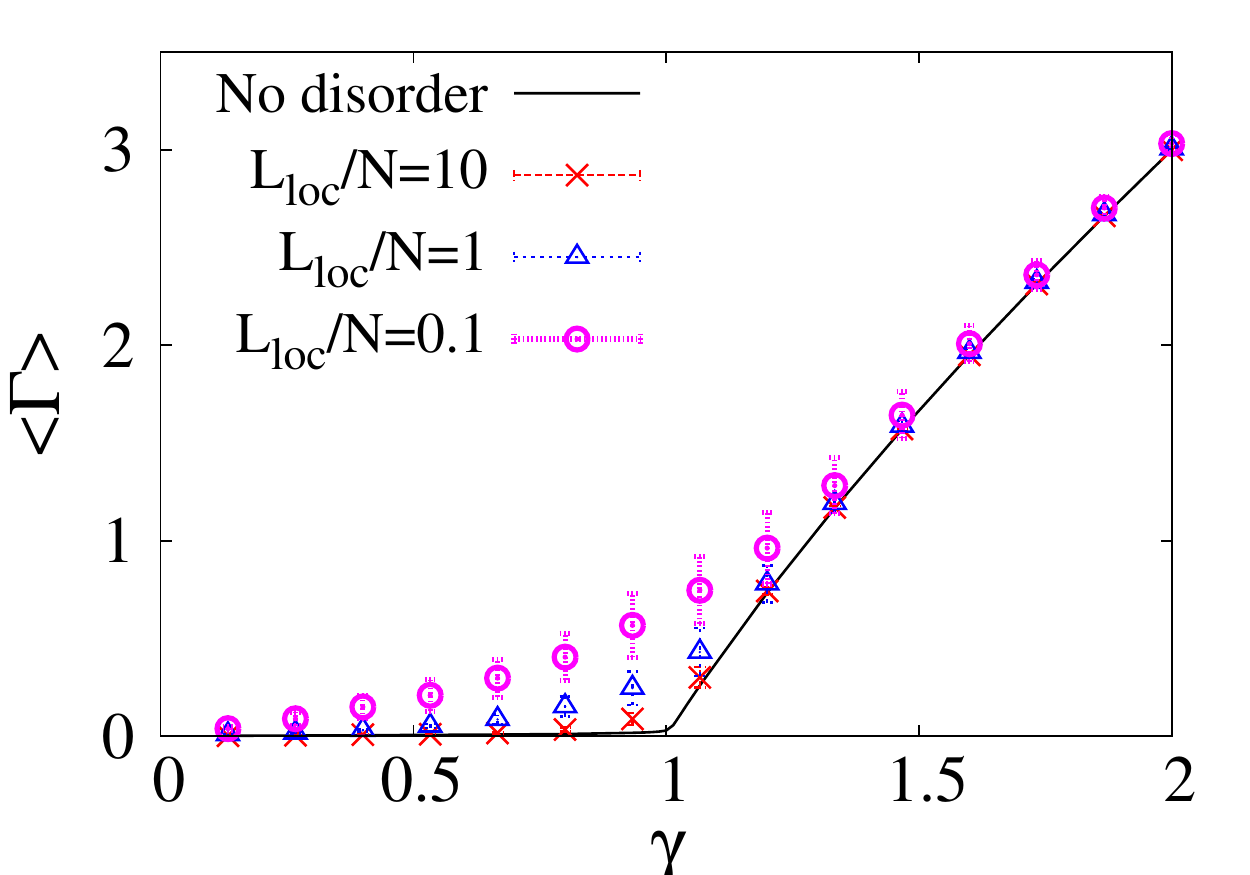}
\end{center}
\caption{\label{figu35} Mean resonance width $\langle\Gamma\rangle$ of the two superradiant
states as a function of the coupling to the continuum $\gamma$ in the ballistic, chaotic, and localized 
regimes for disordered wires of size $N=800$. The energy $E$ was set to zero. Here, 50 
wire realizations were used. The black full line corresponds to the largest eigenvalue of the 
non-Hermitian matrix of (\ref{Eq4}) with $\epsilon_n=0$.}
\end{figure}
%%%%%%%%%%%%%%%%%%%%%%%%%%%%%%%%%%%%%%%%%%%%%%%%%%%%%%%%%%%%%%%%%%%%%%%%%%%%%%%%%%%%%%    

In contrast to the mean width, the standard deviation $\sigma_{\Gamma}$ for individual widths 
depends strongly on the value of the localization length (see error bars in Fig.~\ref{Fig3}). Indeed, 
for the ballistic regime, Fig.~\ref{Fig3}(a), the region with $\gamma\sim 1$ with strong fluctuations 
of $\Gamma$ is quite small as compared with Fig.~\ref{Fig3}(b) and, especially, with Fig.~\ref{Fig3}(c). 
Thus, the fluctuations become larger as the disorder strength is increased, therefore, when the 
localization is stronger. Note that, independently of disorder, at the critical coupling ($\gamma = 1$) 
the fluctuations are so strong that they are of the same size as compared to the mean value of 
$\Gamma$. This fact is characteristic of the phase transitions well studied in statistical 
mechanics.  

The analysis of the error bars in Figs.~\ref{Fig3} shows that they are independent of the system size 
far away from the critical coupling $\gamma=1$. This means that the product $N\sigma_{\Gamma}$ 
is independent of the system size in all regions. Taking into account that neither 
$N\langle \Gamma \rangle$ depends on the system size, one can conclude that the relative fluctuations 
of $\Gamma$ should not vanish in the thermodynamic limit and $\Gamma$ can not be considered as 
a self averaged quantity.

%%%%%%%%%%%%%%%%%%%%%%%%%%%%%%%%%%%%%%%%%%%%%%%%%%%%%%%%%%%%%%%%%%%%%%%%%%%%%%%%%%
\begin{figure*}[!htp]
\includegraphics[width=6cm,height=4cm]{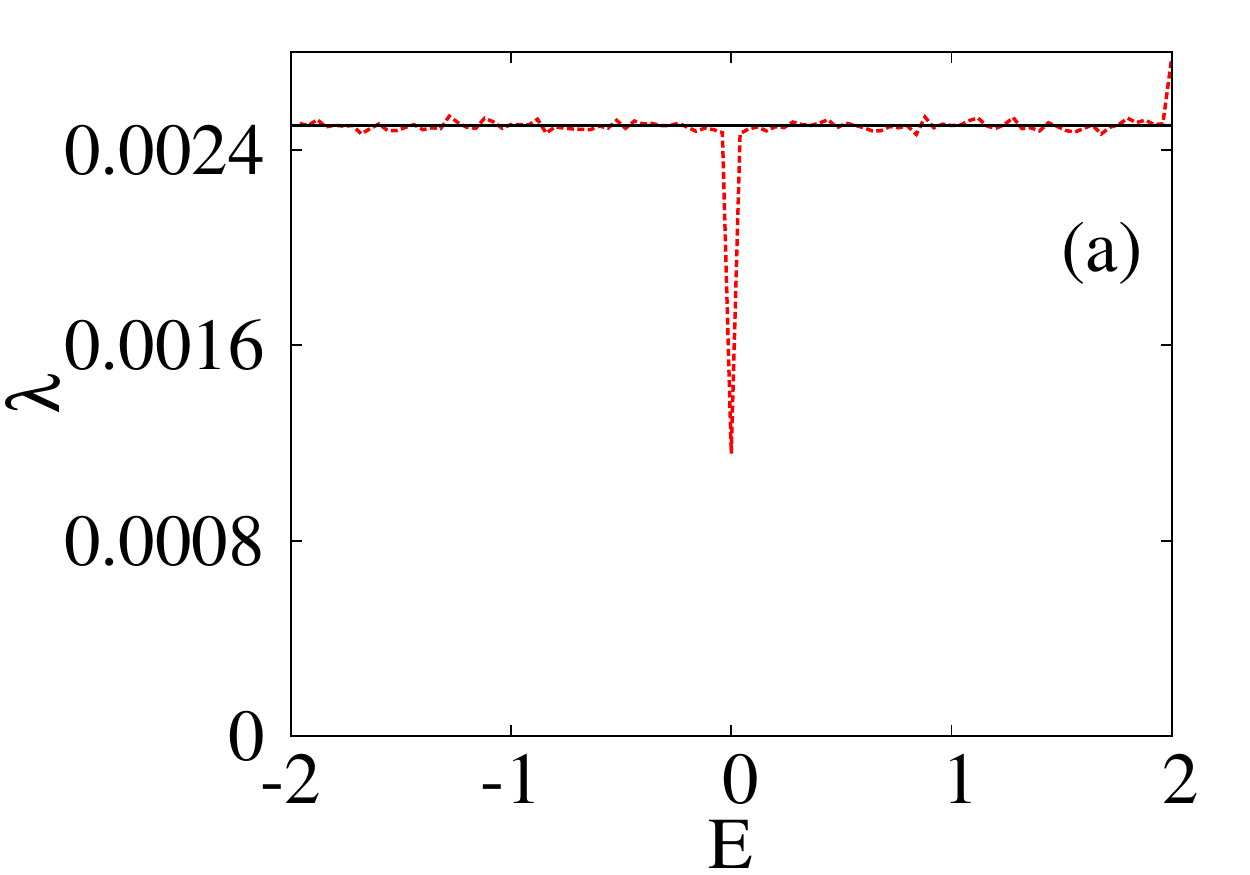}
\includegraphics[width=6cm,height=4cm]{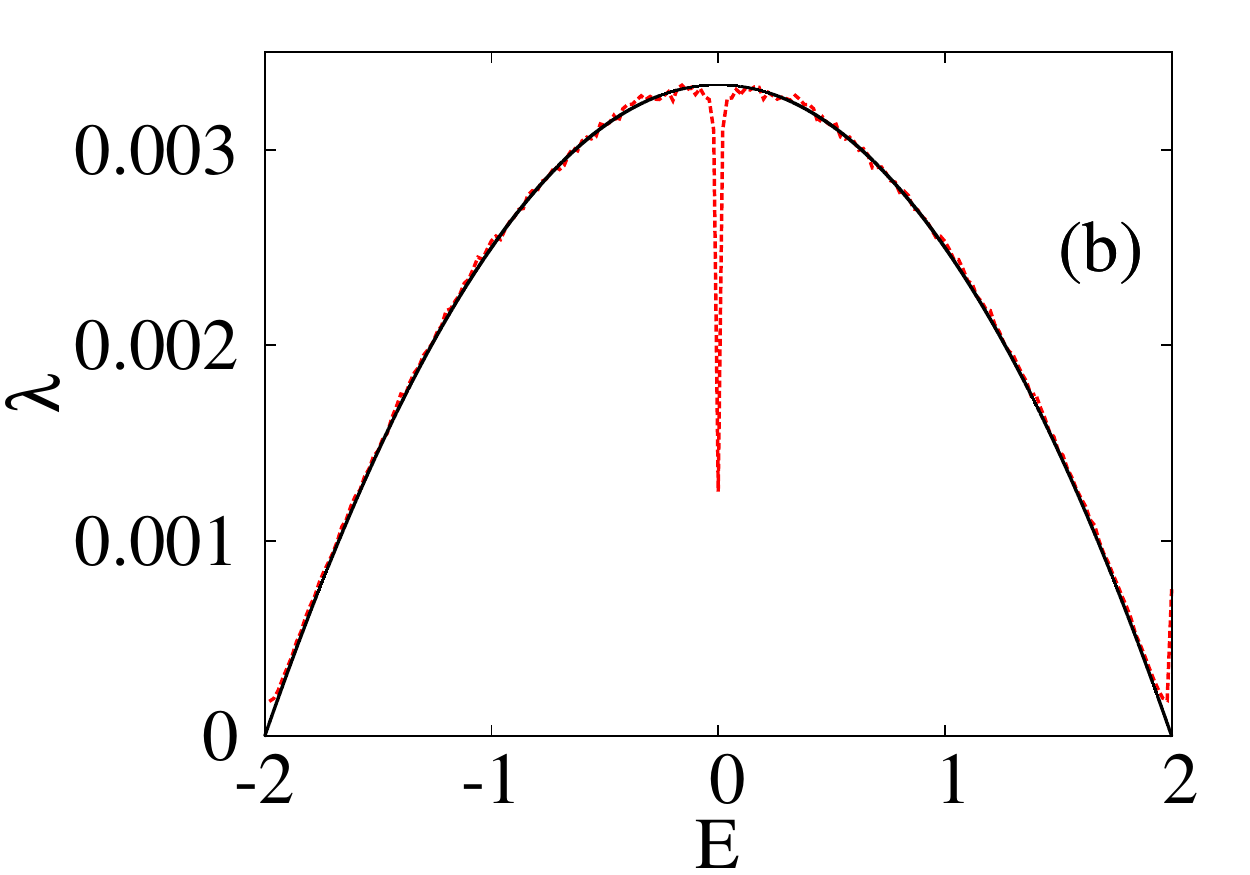}
\includegraphics[width=6cm,height=4cm]{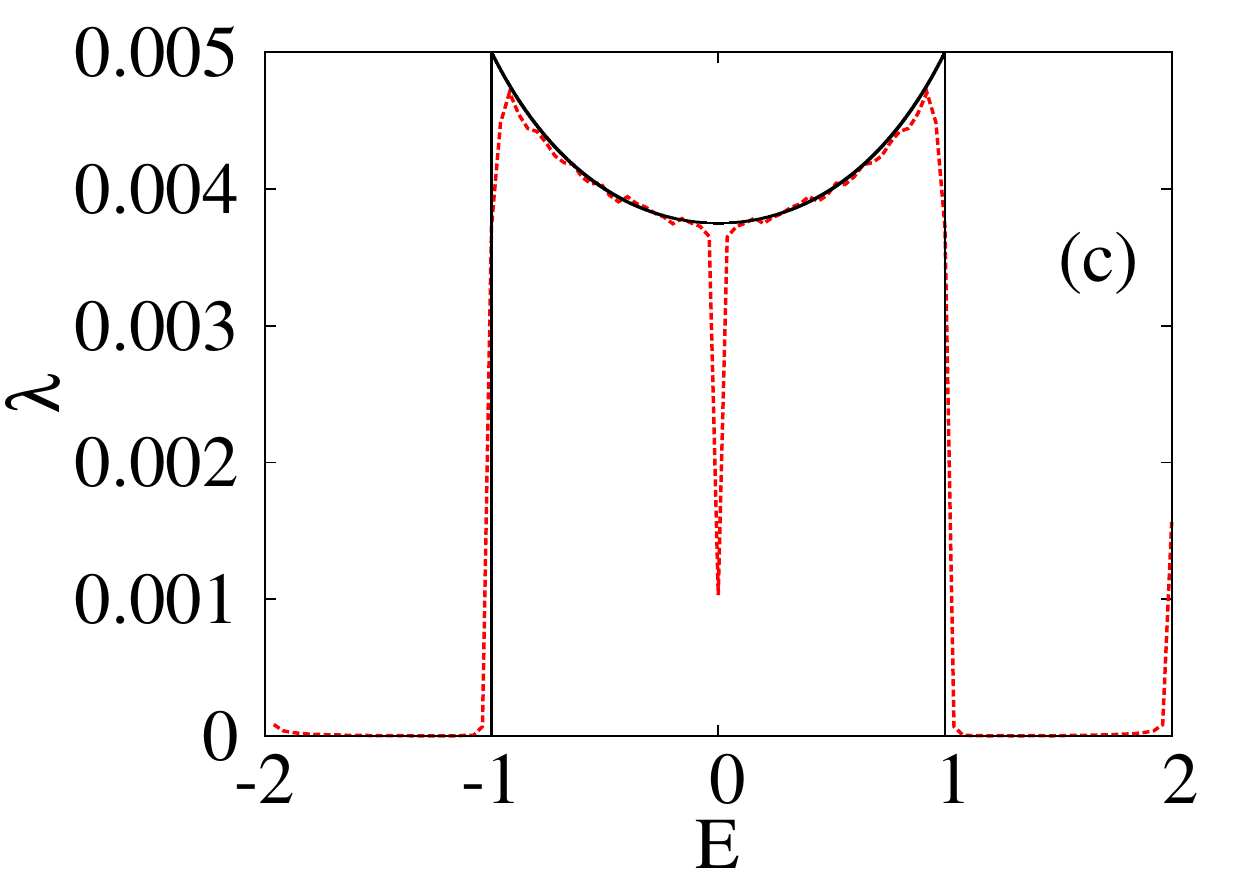}
\includegraphics[width=6cm,height=4cm]{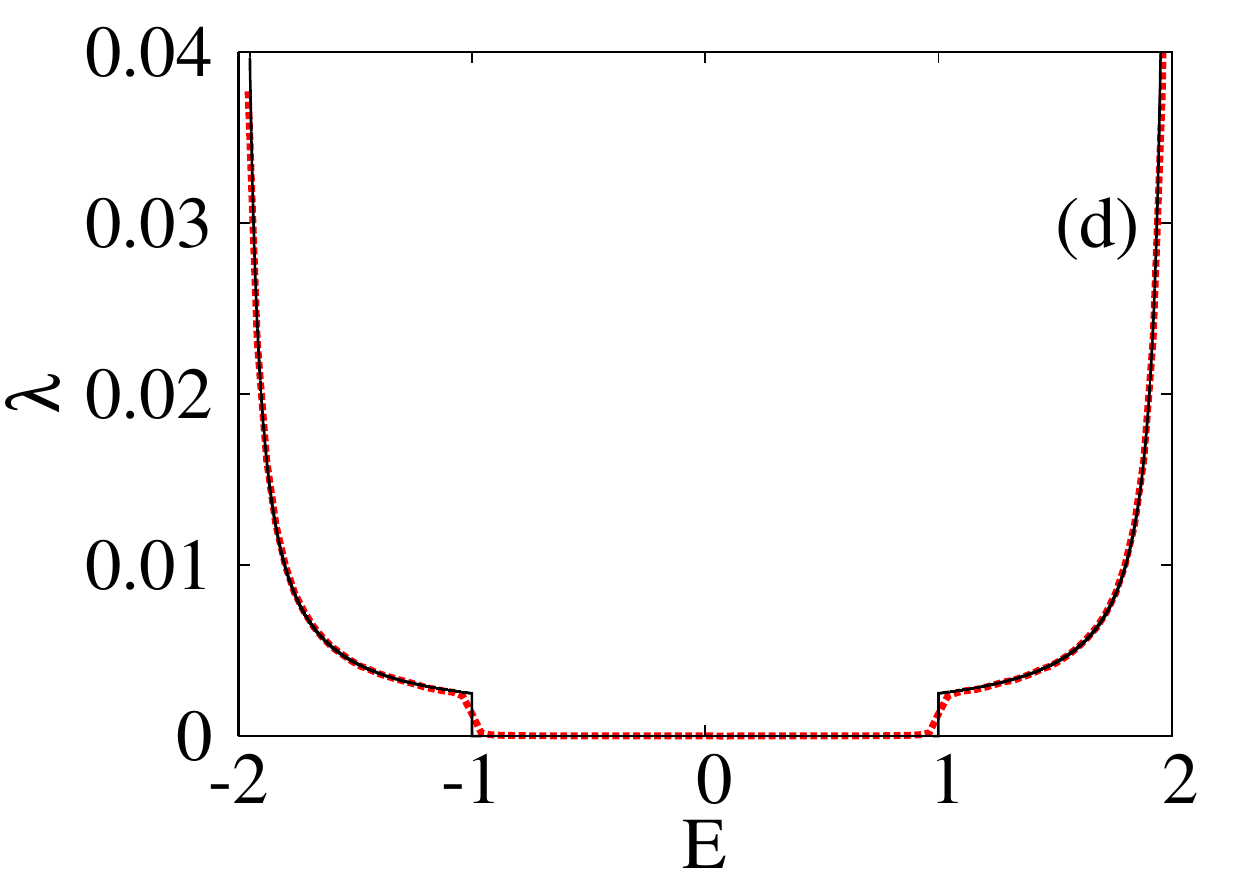}
\caption{\label{Fig10} Inverse localization length $\lambda\equiv L^{-1}_{\mbox{\scriptsize{loc}}}$, 
see Eq.~(\ref{loc1}), as a function of the energy for various correlated disorders. The power spectrum 
$W(\mu)$ of the correlated on-site energies is given by (a) Eq.~(\ref{mu1}), (b) Eq.~(\ref{mu2}), (c) 
Eq.~(\ref{mu3}), and (d) Eq.~(\ref{mu4}). In all cases the disorder intensity was set to $\sigma^2=0.01$. 
Red-dashed curves correspond to numerical data (wires of length $10^8$ were used to compute 
$\lambda$), while the continuous black curves correspond to Eq.~(\ref{loc1}).}
\end{figure*}
%%%%%%%%%%%%%%%%%%%%%%%%%%%%%%%%%%%%%%%%%%%%%%%%%%%%%%%%%%%%%%%%%%%%%%%%%%%

Above, we have focused our attention on the poles corresponding to long-lived states. 
Now, let us look at the poles that correspond to the superradiant states. In Fig.~\ref{figu35} we plot the 
mean width of the $2N_d$ largest-width eigenvalues, where $N_d$ is the number of random 
realizations of the non-Hermitian Hamiltonian matrix (\ref{Eq4}). For $\gamma > 1$ these eigenvalues 
correspond to superradiant states (note that we have two leads attached to each disordered wire). In 
this figure we can clearly see that for large enough coupling to the continuum the mean width 
$\langle \Gamma \rangle$ is practically equal to the largest eigenvalue of the non-Hermitian matrix 
of (\ref{Eq4}) with $\epsilon_n=0$, see the black full line. 
The situation is different in the vicinity of the critical coupling $\gamma=1$, since there the 
disorder plays an important role. In this region it is observed that the shorter the localization length $L_{\mbox{\scriptsize{loc}}}$ the larger the mean width. A similar effect is observed for the 
fluctuations of the widths $\Gamma$ which acquire their largest value close to the critical coupling.

\section{Correlated disorder}

In this Section we consider the case of weak correlated disorder, for which the Thouless expression 
(\ref{loc}) is no more valid and the corresponding localization length gets the form \cite{IKM12}
\begin{eqnarray}
\lambda&\equiv&L^{-1}_{\mbox{\scriptsize{loc}}}=\frac{\sigma^2}{8\sin^2\mu}W(\mu) \ , 
\label{loc1}
\\
W(\mu)&=&1+2\sum^{\infty}_{m=1}K(m)\cos(2\mu m) \ .
\end{eqnarray}
Here, $W(\mu)$ is the power spectrum of on-site energies $\epsilon_n$ and $K(m)$ is the 
normalized binary correlator defined as
\begin{eqnarray}
K(m)=\frac{\langle \epsilon_n \epsilon_{n+m} \rangle}{\sigma^2} \ .
\end{eqnarray}
It is important to stress that Eq.~(\ref{loc1}) is valid for weak disorder, and strong deviations from this 
formula have been found near the band center ($\mu=\pi/2$) and the band edges ($\mu=0,\pi$), see 
details in Ref.~\cite{IKM12}. Notice that for uncorrelated disorder we have $W(\mu)=1$ and, therefore, 
Eq.~(\ref{loc1}) reduces to Eq.~(\ref{loc}). Here, $\lambda\equiv L^{-1}_{\mbox{\scriptsize{loc}}}$ is also 
known as the Lyapunov exponent. In what follows, we consider various types of correlations imposed to
disordered potentials. 

\subsection{Constant localization length}

In comparison with Eq.~(\ref{loc}), expression (\ref{loc1}) contains the additional energy-dependent 
term $W(\mu)$. This fact allows one to impose specific correlations for a given energy dependence 
of the localization length along the energy band, that is of great interest for various applications 
\cite{IKM12}. Let us start with the simplest, however, non-trivial case of disorder for which the localization 
length does not depend on energy. In this case the power spectrum takes the form,
\begin{eqnarray}
W(\mu)=2\sin^2\mu \ , 
\label{mu1}
\end{eqnarray}
therefore, the inverse of the localization length is constant: $L^{-1}_{\mbox{\scriptsize{loc}}}=\sigma^2/4$. 
The corresponding binary correlator is given by 
\begin{equation}
K(m)=\delta_{m,0}-\frac{1}{2}\delta_{\left| m\right|,1} \ .
\end{equation}
This correlator has only three components, $K(0)=1$ and $K(\pm1)=-1/2$, the other components vanish, 
thus the correlations are short-range. Figure.~\ref{Fig10}(a) shows an excellent 
agreement between the numerically obtained $L^{-1}_{\mbox{\scriptsize{loc}}}$ and Eq.~(\ref{loc}), except 
in the vicinity of the band center where a clear resonant behavior emerges. The region near the band 
center has been studied in detail (see Ref.~\cite{IKM12} and references therein), however in this paper we 
are interested in the generic properties of the localization length, therefore we focus on the energies far enough 
from the band center and band edges. 

\begin{figure*}[!htp]
\includegraphics[width=5.5cm,height=4cm]{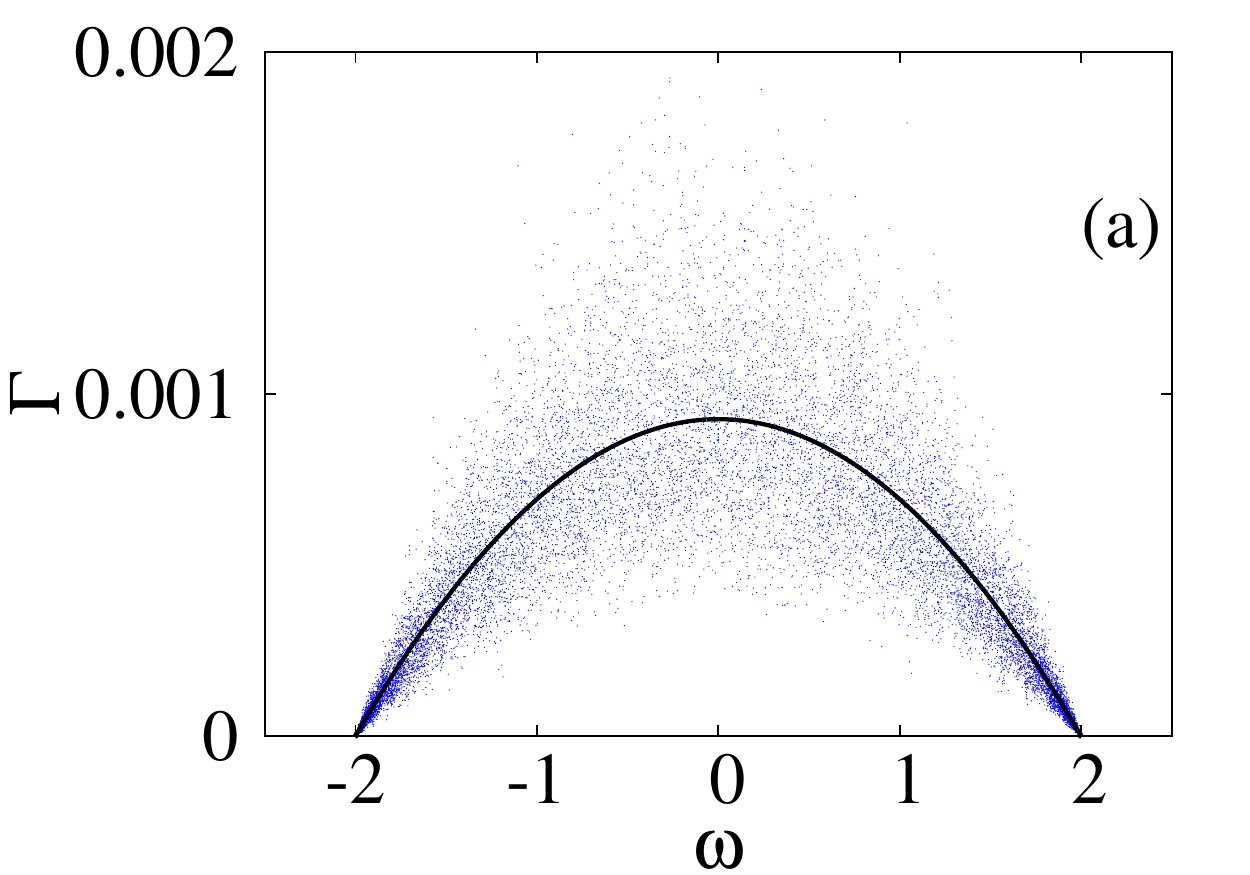}
\includegraphics[width=5.5cm,height=4cm]{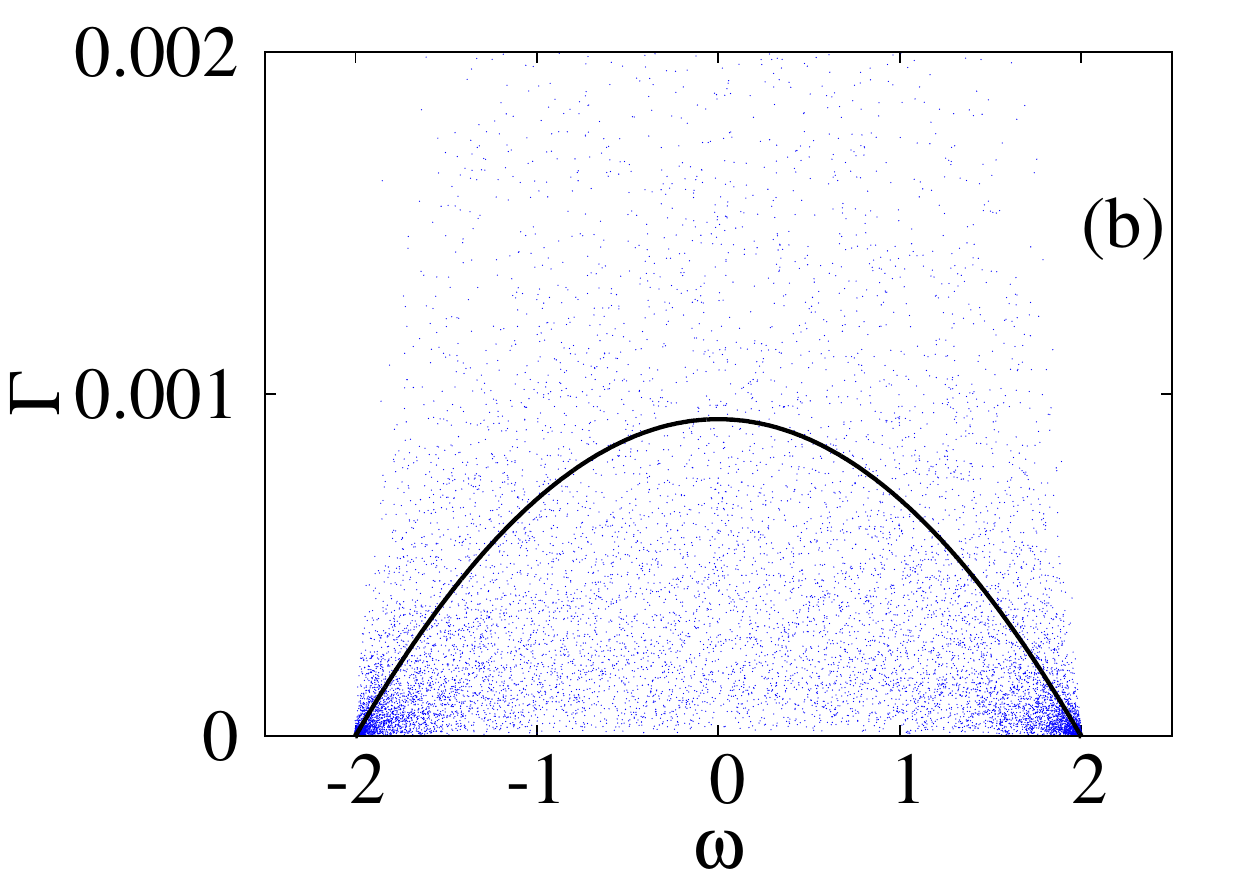}
\includegraphics[width=5.5cm,height=4cm]{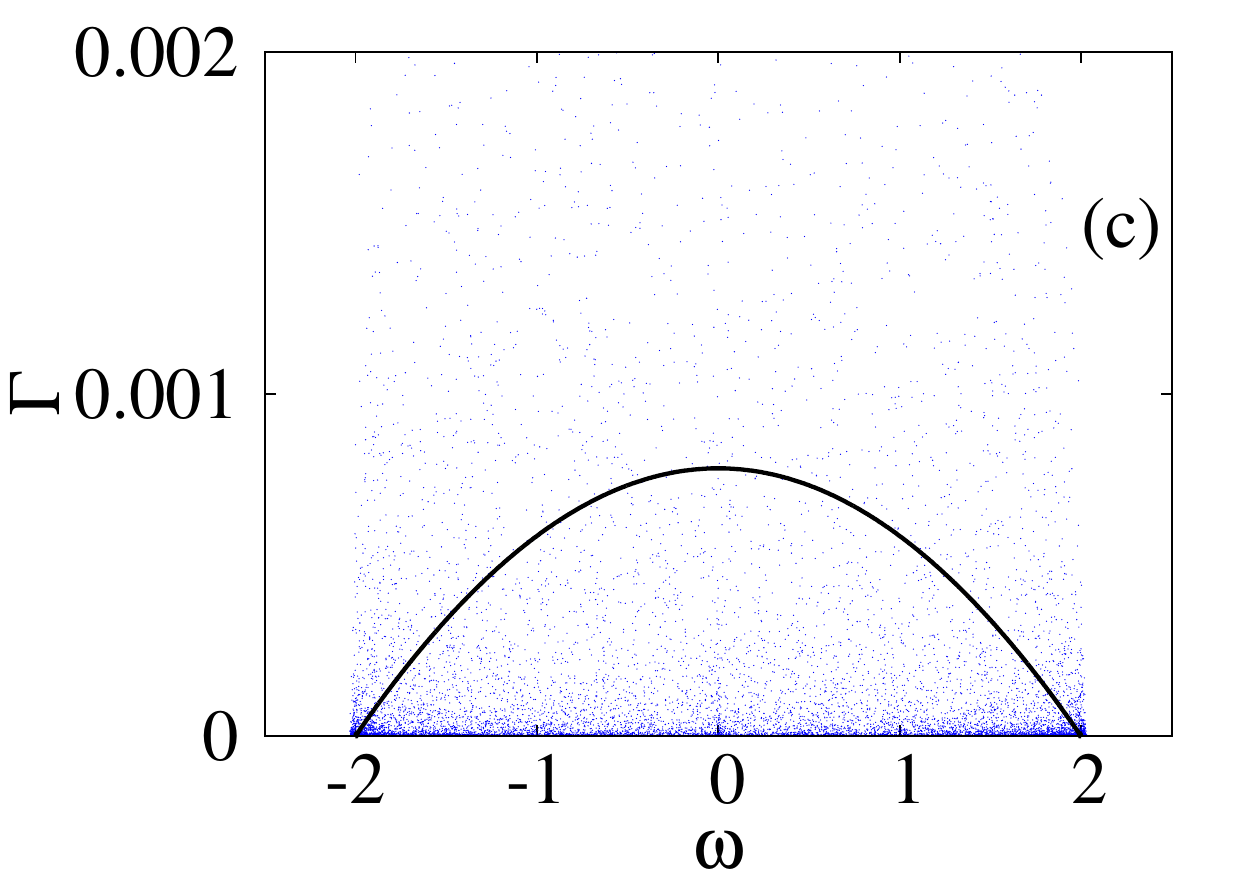}
\includegraphics[width=5.5cm,height=4cm]{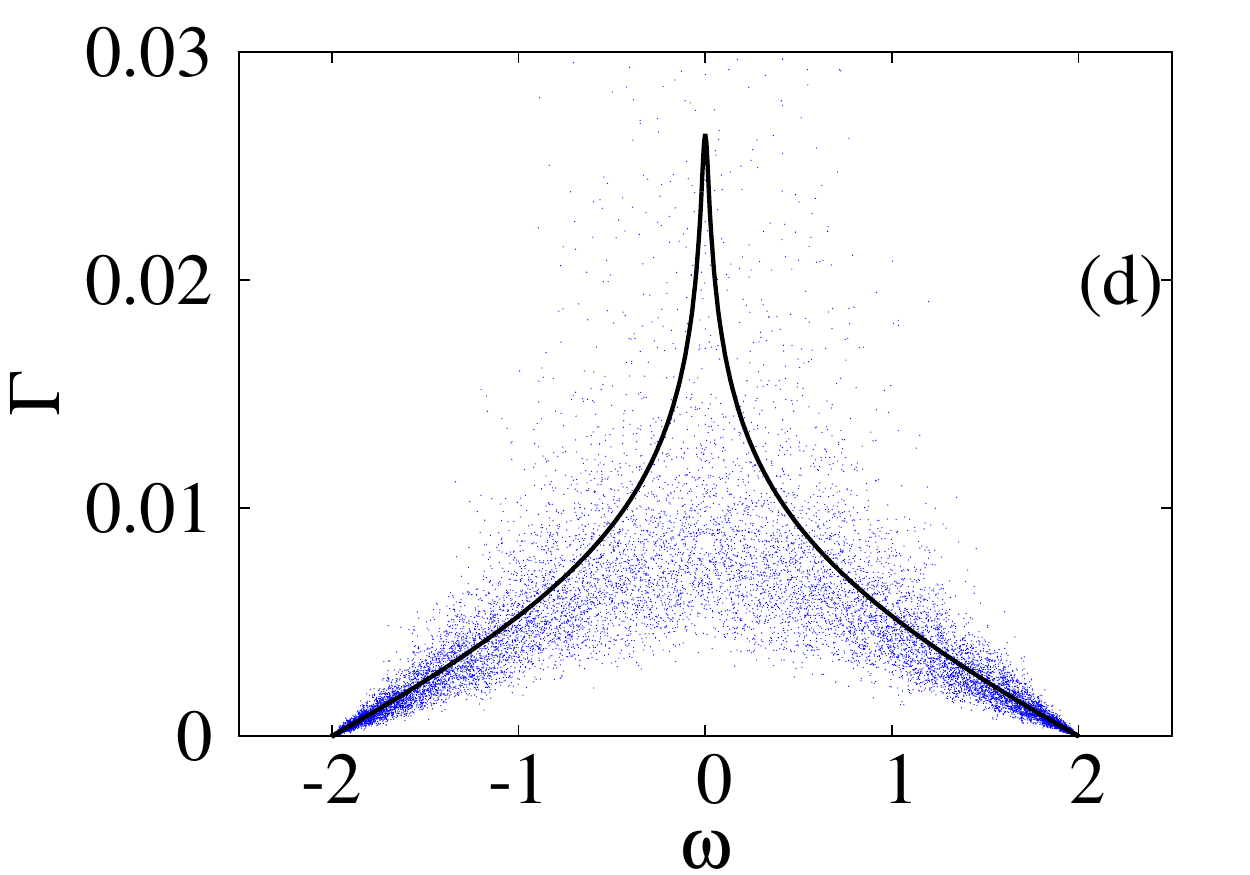}
\includegraphics[width=5.5cm,height=4cm]{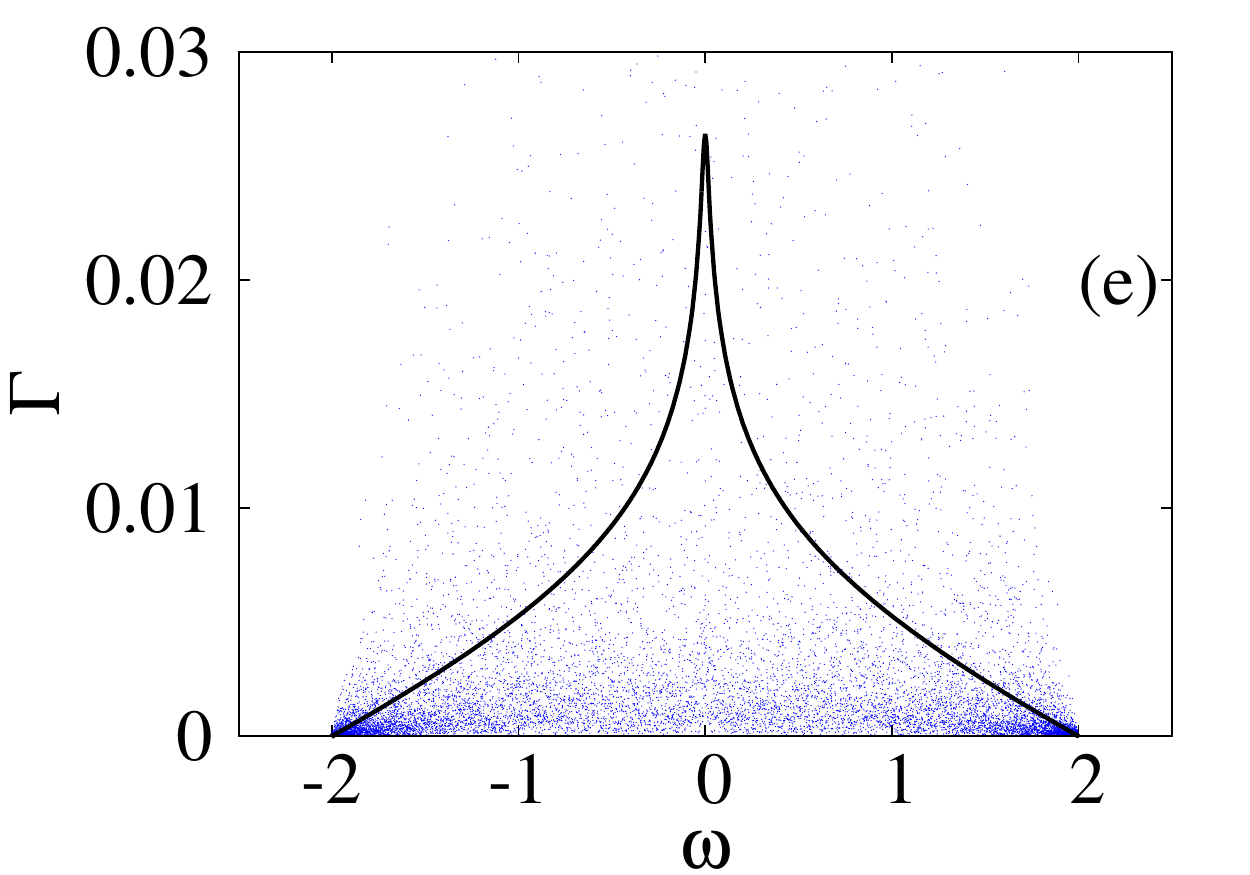}
\includegraphics[width=5.5cm,height=4cm]{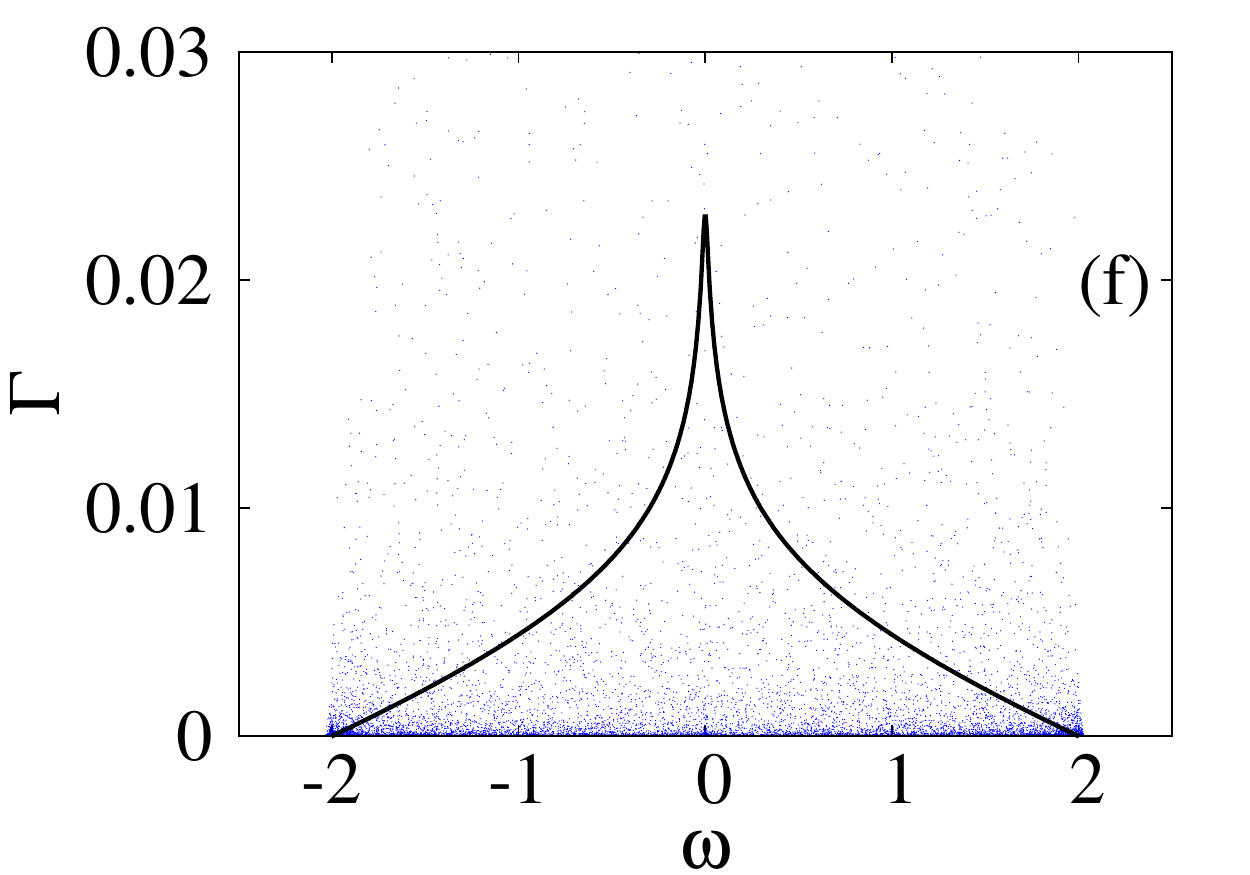}
\caption{\label{Fig6} Imaginary vs.~real part of the $S$-matrix poles $\Omega$ [see Eq.~(\ref{poles})] for 1D 
wires with correlated 
disorder defined by the power spectrum (\ref{mu1}). Wires of length $N=865$ are coupled to the continuum 
with strengths $\gamma=0.1$ (upper panels) and $\gamma=1$ (lower panels). The disorder strength was 
set to $\sigma^2=0.001$, $L_{\mbox{\scriptsize{loc}}}/N=10$ (left panels); 
$\sigma^2=0.01$, $L_{\mbox{\scriptsize{loc}}}/N \approx 1$ (middle panels); and 
$\sigma^2=0.1$, $L_{\mbox{\scriptsize{loc}}}/N=0.1$ (right panels). 
Black curves represent the corresponding non-disordered wires. Here, 50 wire realizations were used.}
\end{figure*}

Correspondingly, in Fig.~\ref{Fig6} we present the distribution of the $S$-matrix poles in the complex plane 
for the correlated disorder with the power spectrum of Eq.~(\ref{mu1}). The disorder increases from left to 
write panels; specifically, left panels correspond to weak disorder while right panels to strong disorder. 
The upper panels show the pole distribution for weak coupling ($\gamma \ll 1$) and the low panels are 
given for strong coupling ($\gamma \approx 1$).

For the ballistic regime (left panels; weak disorder) the poles are distributed around the curves 
corresponding to zero disorder (shown in the figure as black curves). In this regime in which all 
eigenstates are extended and close to plane waves, the embedded correlations in the potential 
do not modify strongly the distribution of poles. In contrast, in the chaotic regime (central panels) 
the poles are strongly scattered in the complex plane. Whereas for the localized regime (right 
panels), we observe the distribution of poles close to that occurring for the uncorrelated disorder. 
In this case, the effect of the coupling to the continuum is strongly reduced 
due to the small amount of eigenstates that are connected to the leads. In this case the poles 
are located quite close to real axis, therefore, the system should be treated as strongly 
isolated from the continuum. One can conclude that the most significant differences (between 
uncorrelated and correlated disordered wires with constant localization length) appears when the 
eigenstates of the Hermitian $H$ are both delocalized and chaotic. This situation occurs when 
the localization length is of the order of the system size (see middle panels in Fig.~\ref{Fig6}).

%%%%%%%%%%%%%%%%%%%%%%%%%%%%%%%%%%%%%%%%%%%%%%%%%%%%%%%%%%%%%%%%%%%%%%%%%%%%%%%%%%%%%%%%%%%
%%%%%%%%%%%%%%%%%%%%%%%%%%%%%%%%%%%%%%%%%%%%%%%%%%%%%%%%%%%%%%%%%%%%%%%%%%%%%%
%%%%%%%%%%%%%%%%%%%%%%%%%%%%%%%%%%%%%%%%%%%%%%%%%%%%%%%%%%%%%%%%%%%%%%%%%%%%%%
\begin{figure*}[!htp]
\includegraphics[width=5.5cm,height=4cm]{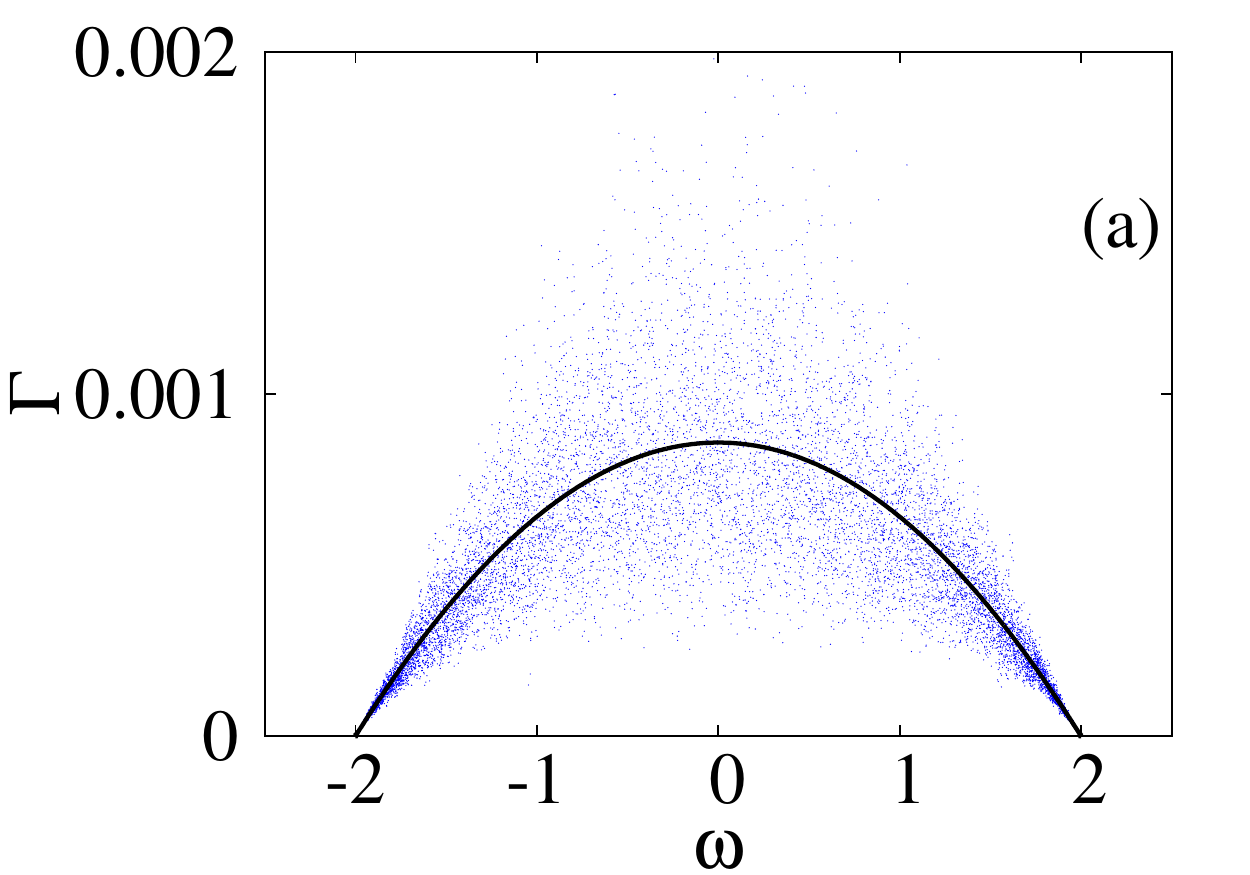}
\includegraphics[width=5.5cm,height=4cm]{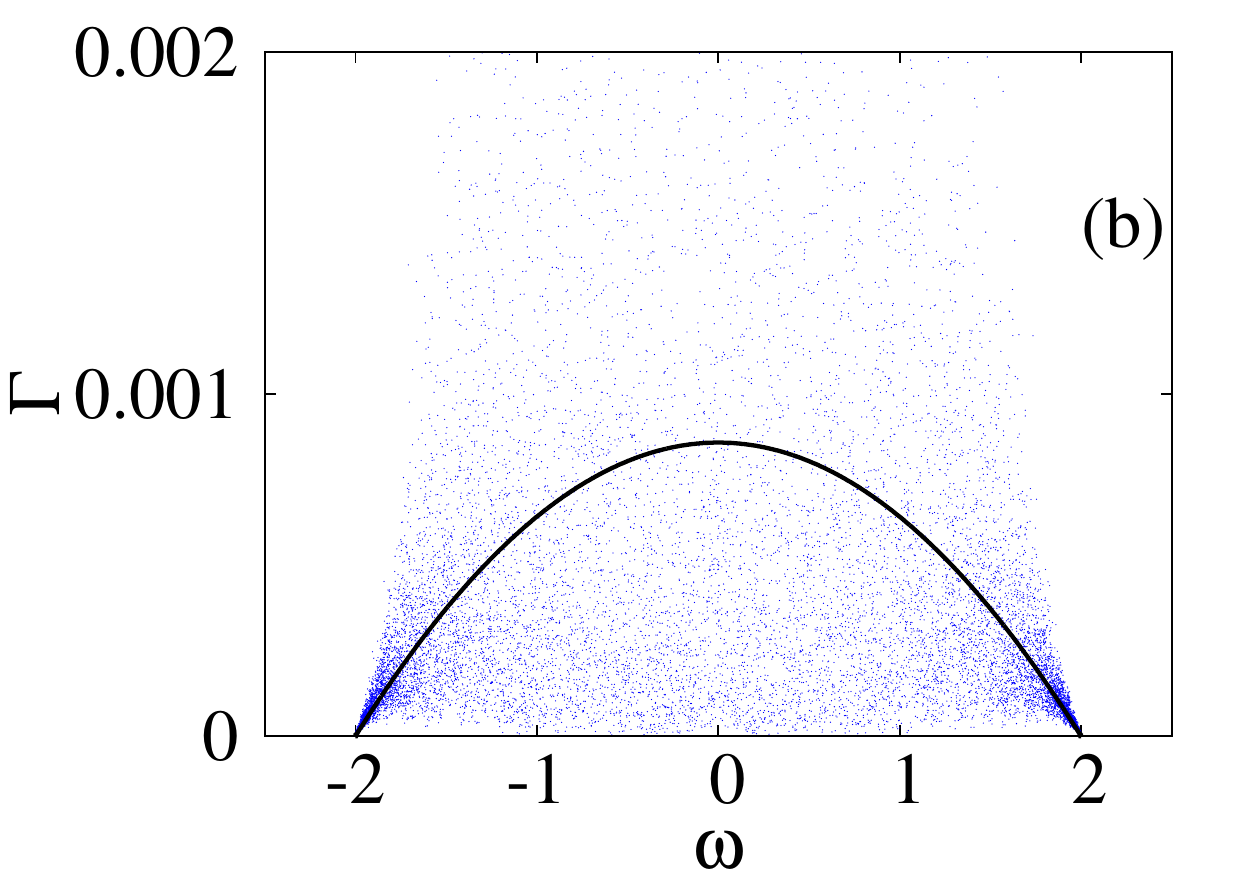}
\includegraphics[width=5.5cm,height=4cm]{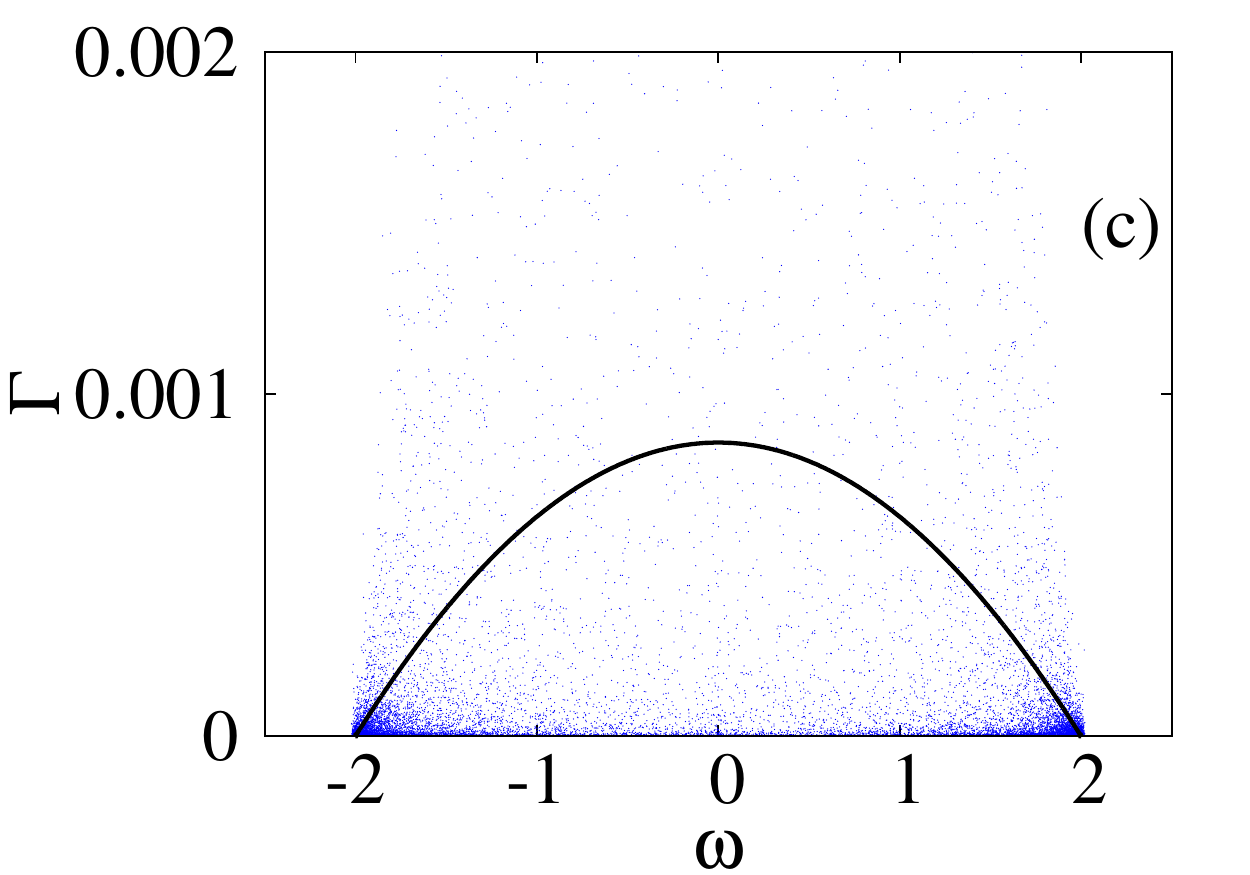}
\includegraphics[width=5.5cm,height=4cm]{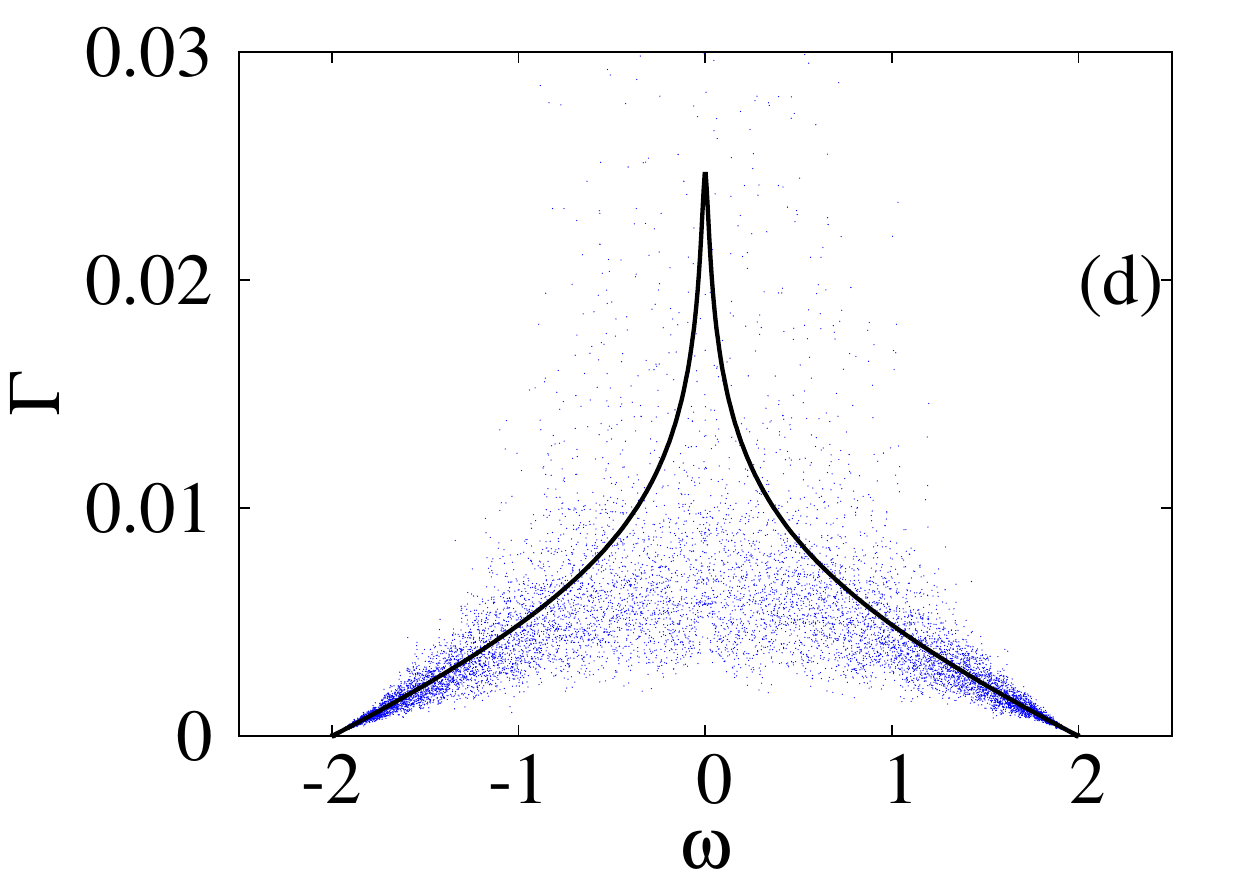}
\includegraphics[width=5.5cm,height=4cm]{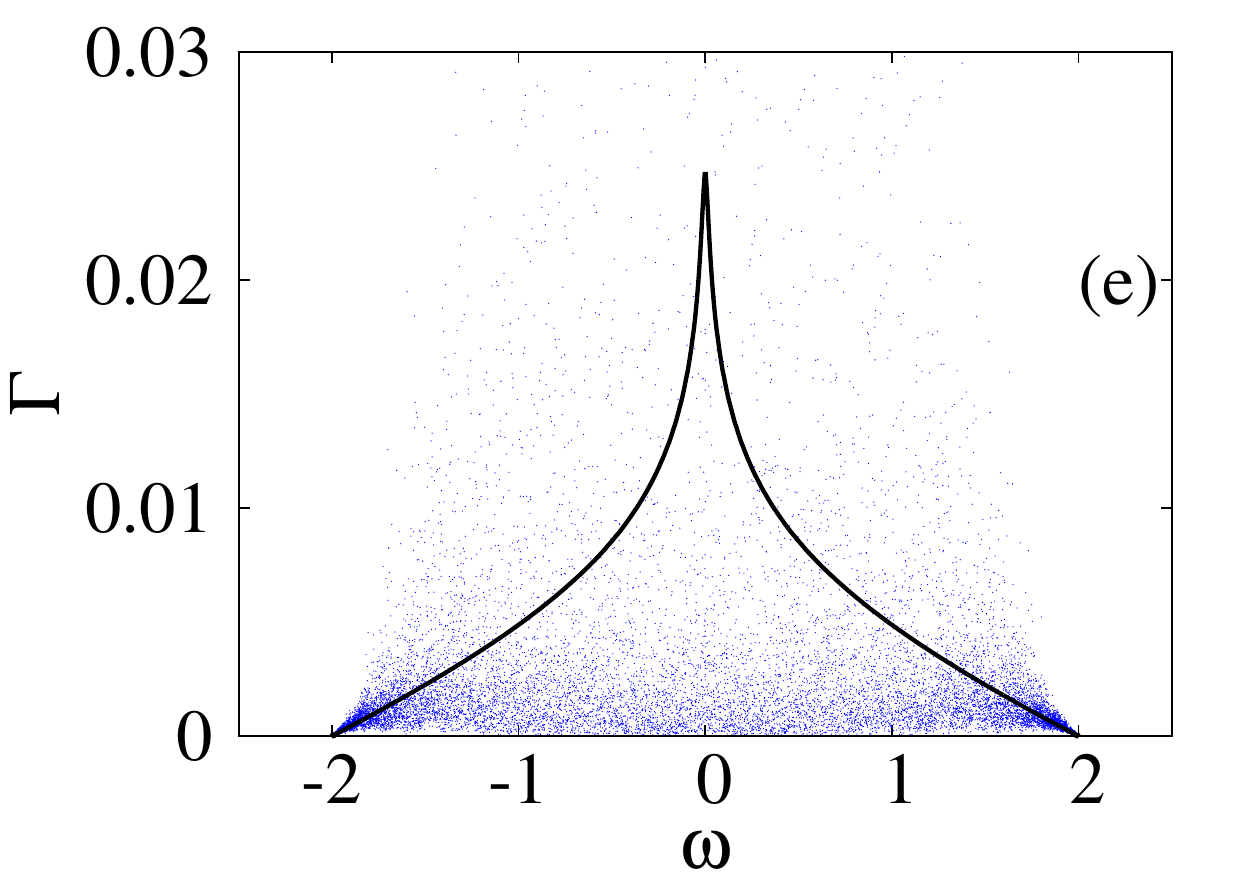}
\includegraphics[width=5.5cm,height=4cm]{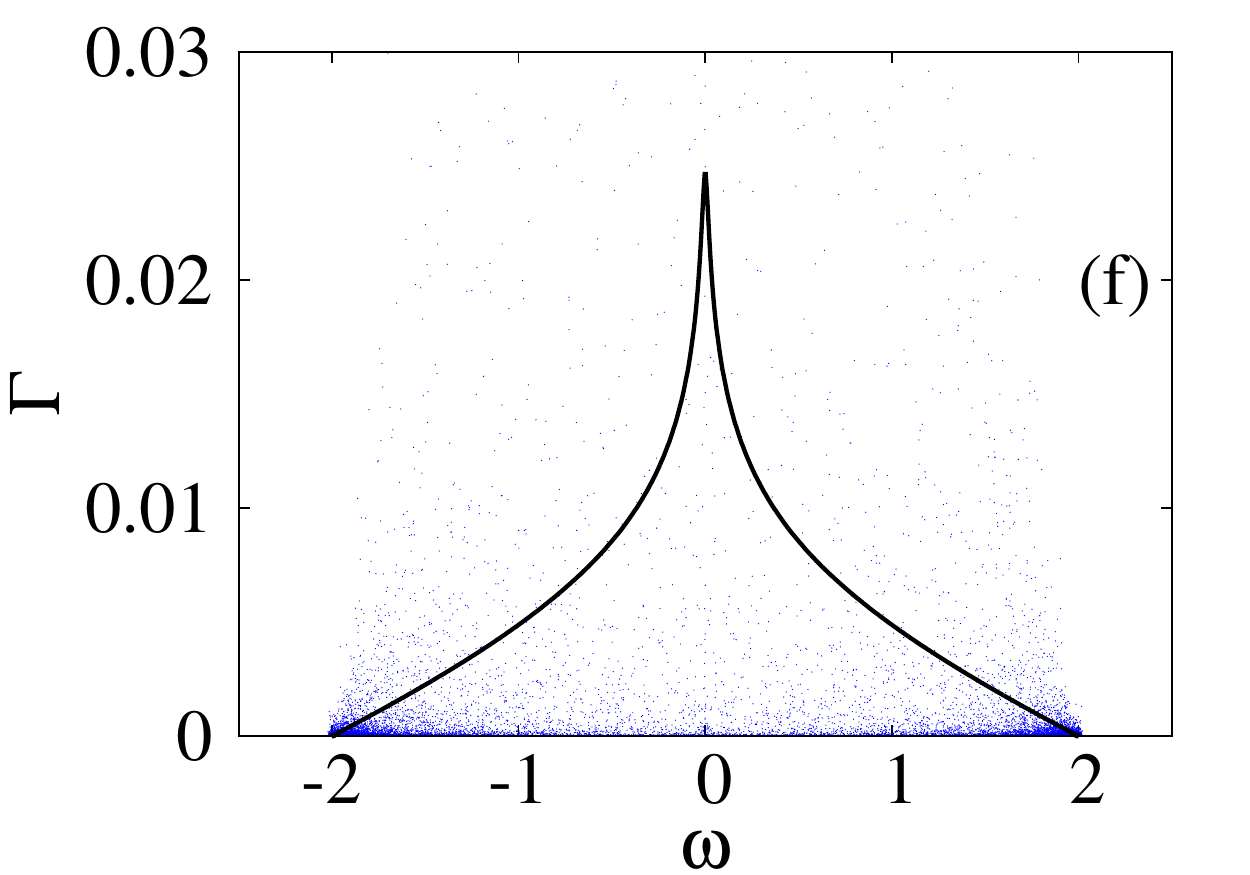}
\caption{\label{Fig7} Imaginary vs.~real part of the $S$-matrix poles $\Omega$ [see Eq.~(\ref{poles})] for 1D 
wires with correlated 
disorder defined by the power spectrum (\ref{mu2}). Wires of length $N=934$ are coupled to the continuum 
with strengths $\gamma=0.1$ (upper panels) and $\gamma=1$ (lower panels). The disorder strength was set to 
$\sigma^2=0.001$, $L_{\mbox{\scriptsize{loc}}}/N=10$ (left panels); 
$\sigma^2=0.01$, $L_{\mbox{\scriptsize{loc}}}/N \approx 1$ (middle panels); and 
$\sigma^2=0.1$, $L_{\mbox{\scriptsize{loc}}}/N=0.1$ (right panels). 
Black curves represent the corresponding non-disordered wires. Here, 50 wire realizations were used.}
\end{figure*}
%%%%%%%%%%%%%%%%%%%%%%%%%%%%%%%%%%%%%%%%%%%%%%%%%%%%%%%%%%%%%%%%%%%%%%%%%%%%%%%%

\subsection{Inverse localization length proportional to $\sin^2\mu$}

In the uncorrelated disorder case we have 
$L^{-1}_{\mbox{\scriptsize{loc}}}(\mu) \propto \sin^{-2}\mu$, see Eq.~(\ref{loc}); that is, 
$L^{-1}_{\mbox{\scriptsize{loc}}}$ has a minimum at $E=0$ and diverges at the energy band 
edges $E = \pm 2$. Using correlated disorder it is possible to {\it invert} this behavior: i.e., making 
$L^{-1}_{\mbox{\scriptsize{loc}}}$ to have a maximum at $E=0$ and make it vanishing at $E = \pm 2$. 
To this end we use the correlations defined by the power spectrum
\begin{eqnarray}
W(\mu)=\frac{8}{3}\sin^4 \mu \ ,
\label{mu2}
\end{eqnarray}
so that $L^{-1}_{\mbox{\scriptsize{loc}}}= \sigma^2\sin^2\mu/3$. Here Eq.~(\ref{mu2}) corresponds to the 
following binary correlator:
\begin{eqnarray}
K(m)=\delta_{m,0}-\frac{2}{3}\delta_{\left|m\right|,1}+\frac{1}{6}\delta_{\left|m\right|,2} \ .
\end{eqnarray}
Thus, the non-vanishing components of $K(m)$ corresponds to $m=0$, $m=\pm 1$, and $m=\pm 2$. 

The study of our model with the above correlations manifests a good correspondence between the 
analytical formula for the localization length (\ref{loc1}) and numerical data as shown in Fig.~\ref{Fig10}(b). 
Note that a narrow resonance at the band center persists, indicating a typical deviation from Eq.~(\ref{loc1}). 
The corresponding pole distributions are shown in Fig.~\ref{Fig7}. One can see that the inclusion of 
this kind of correlated disorder does not change too much the ballistic regime picture. Indeed, the poles in 
Fig.~\ref{Fig7} (left panels) are distributed  pretty much the same as for the non-disordered case. As the 
localization length increases and reaches the chaotic regime (middle panels), the gap at the band center 
(clearly seen in the case of uncorrelated disorder) is now practically negligible. Finally, for the localized 
regime (right panels) most of the poles are located close to the real axis. A distinctive feature for this type 
of correlations is that most of the poles are concentrated in the vicinity of the band edges ($\omega=\pm 2$).

\subsection{Mobility edges I}

Now we turn our attention to an interesting situation for which the power spectrum has the form
\begin{eqnarray}
     W(E)= 
         \left\{
                \begin{array}{ll}
                   \pi/(\pi-2\mu_1) \ , &  \quad -E_1<E<E_1 \ , \\
                     0 \ , & \quad \mbox{otherwise} \ .
                \end{array} 
         \right.
\label{mu3}
\end{eqnarray}
In this case the localization length in the corresponding isolated system (for $\gamma=0$) is strongly 
suppressed inside the energy interval $[-E_1,E_1]$ and enhanced outside this interval. Thus, the critical 
values $\pm E_1$ can be treated as the mobility edges (see details 
and discussion in Ref.~\cite{IKM12}). Note that the value of $\mu_1$ is simply determined by the dispersion 
relation $E_1=2\cos \mu_1$. In our numerical simulation, we set $E_1=1$. The power spectrum (\ref{mu3}) 
corresponds to the following binary correlator:
\begin{eqnarray}
K(m)=\frac{1}{m(\pi-2\mu_1)}\sin(2m\mu_1) \ .
\label{eq3}
\end{eqnarray}
This correlator exhibits a power law decay typical of long-range correlated disorder. 

The prediction of an effective delocalization transition, occurring in the first order approximation with 
respect to weak disorder, is corroborated in Fig.~\ref{Fig10}(c). Here, a good agreement between 
numerical data and the analytical expression (\ref{loc1}) is clearly seen. A strong discrepancy emerges 
around the band center only, where the influence of the resonant behavior cannot be neglected.

\begin{figure*}[!htp]
\includegraphics[width=5.5cm,height=4cm]{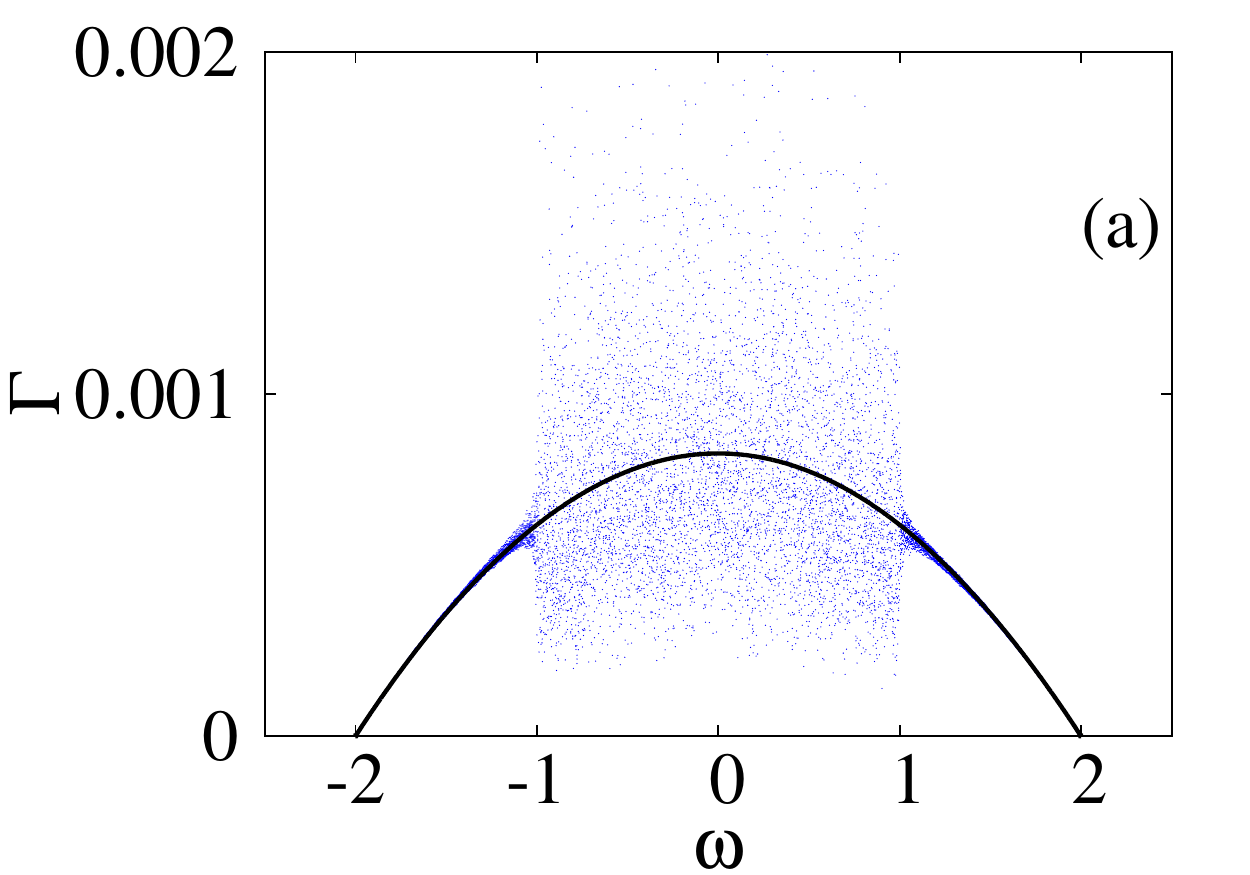}
\includegraphics[width=5.5cm,height=4cm]{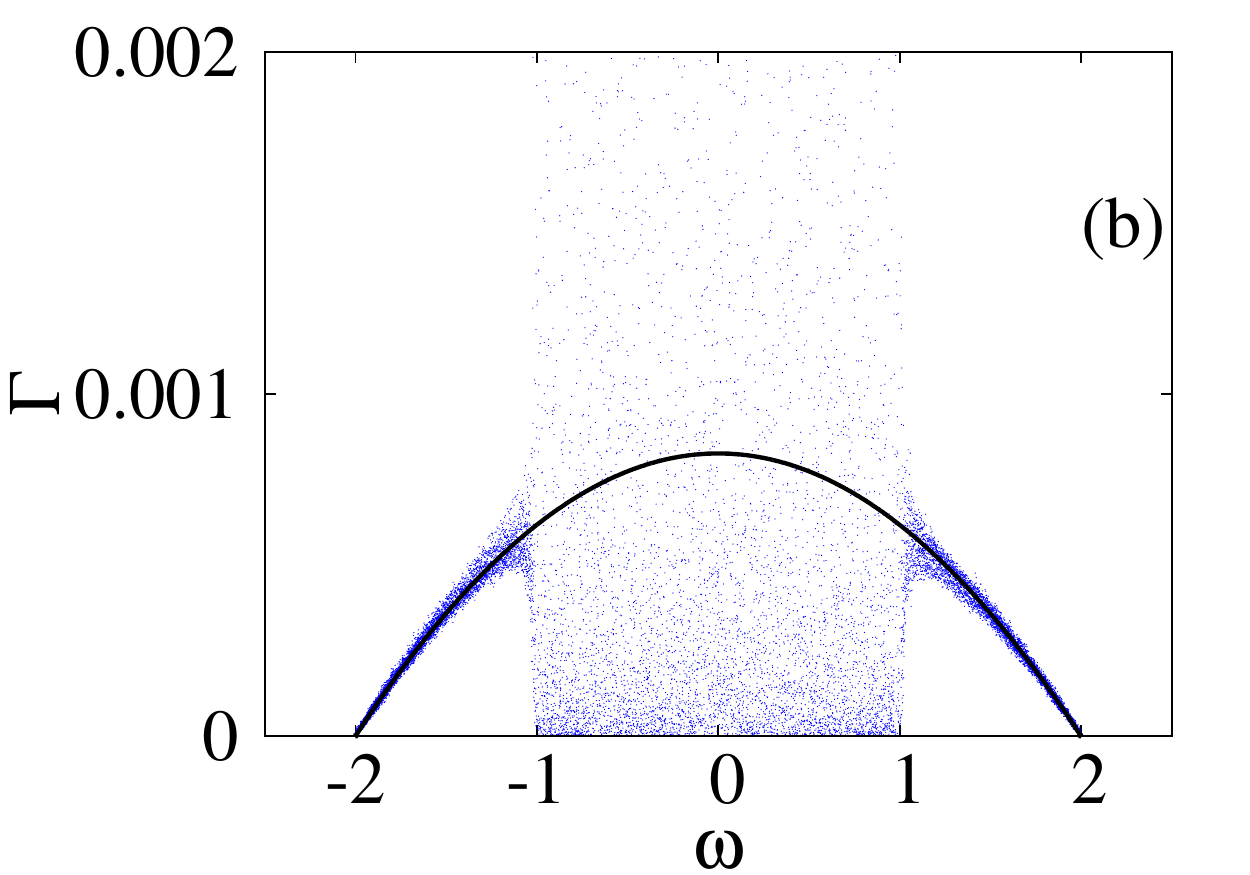}
\includegraphics[width=5.5cm,height=4cm]{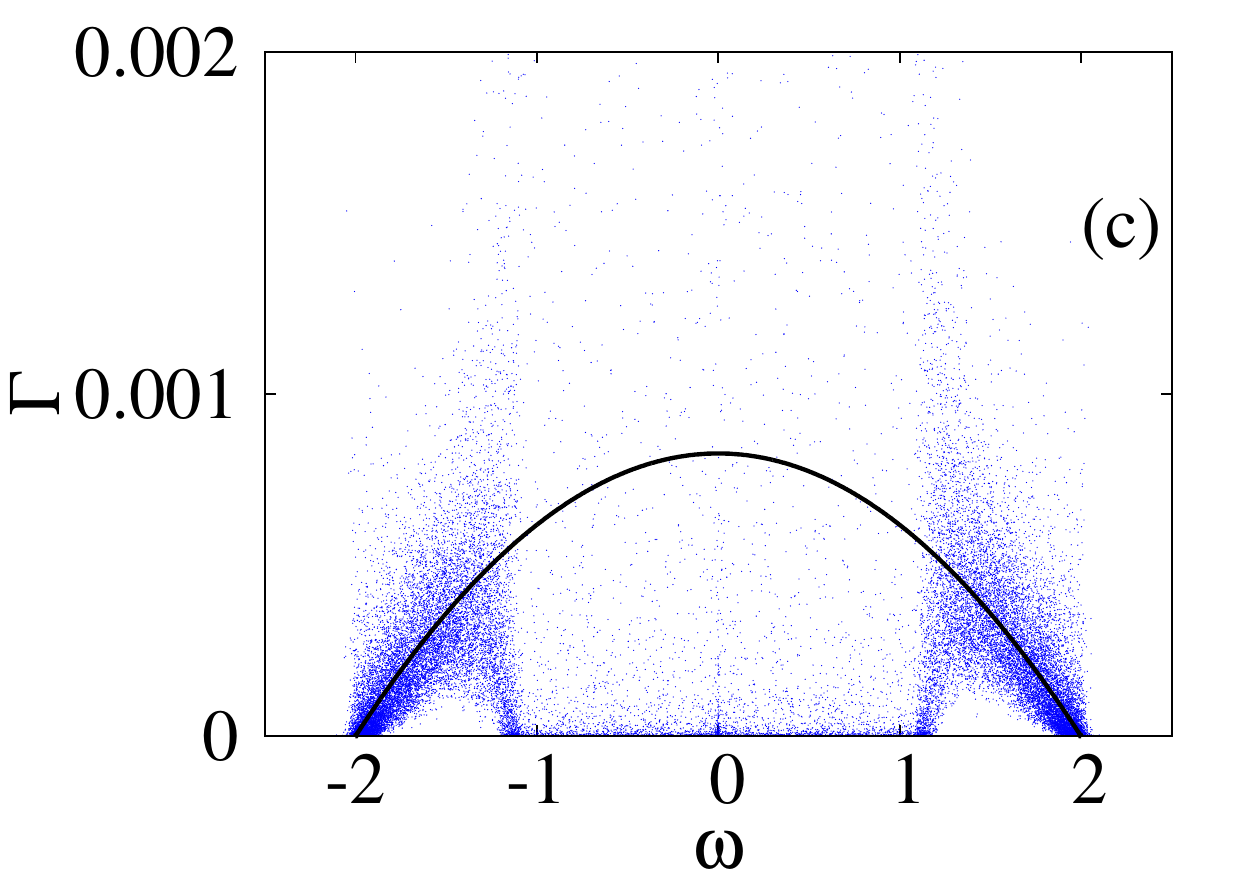}
\includegraphics[width=5.5cm,height=4cm]{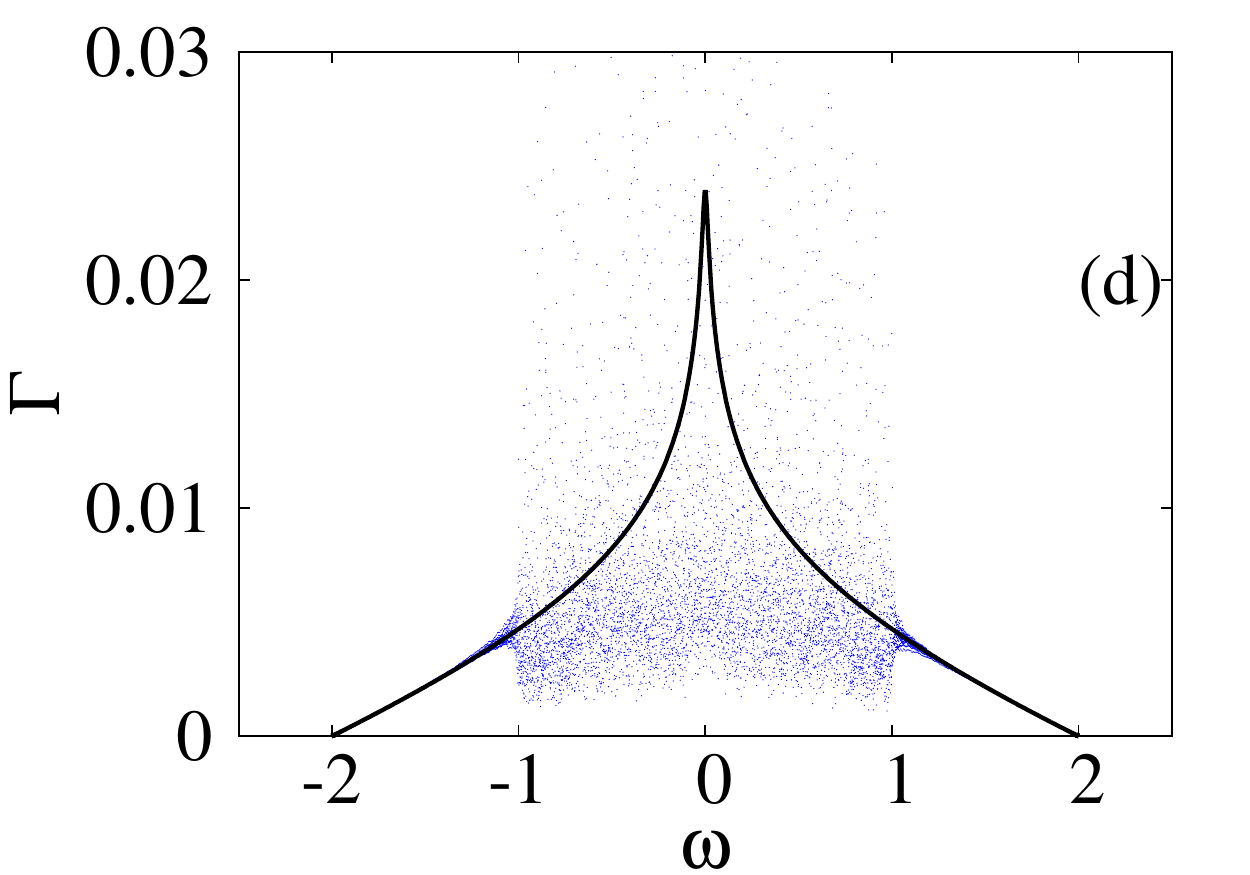}
\includegraphics[width=5.5cm,height=4cm]{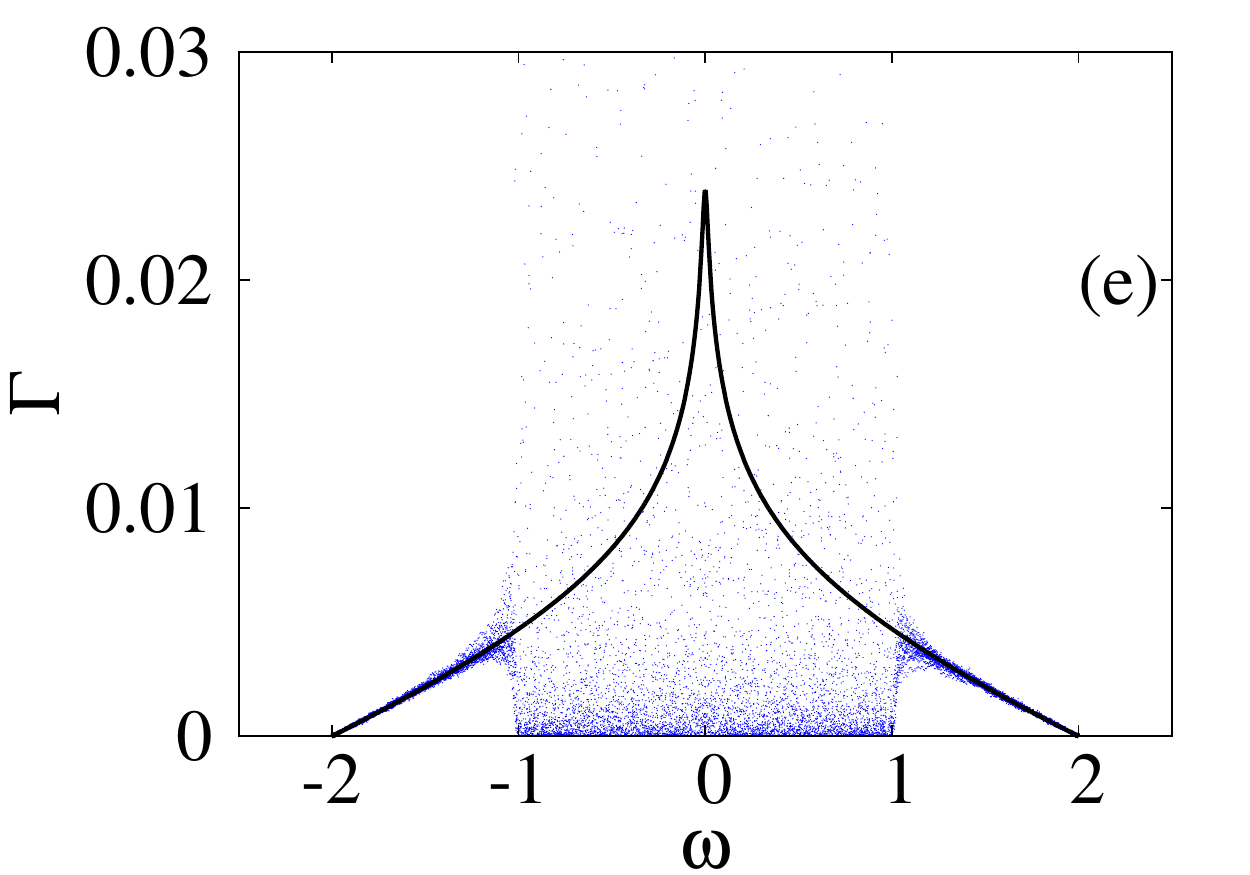}
\includegraphics[width=5.5cm,height=4cm]{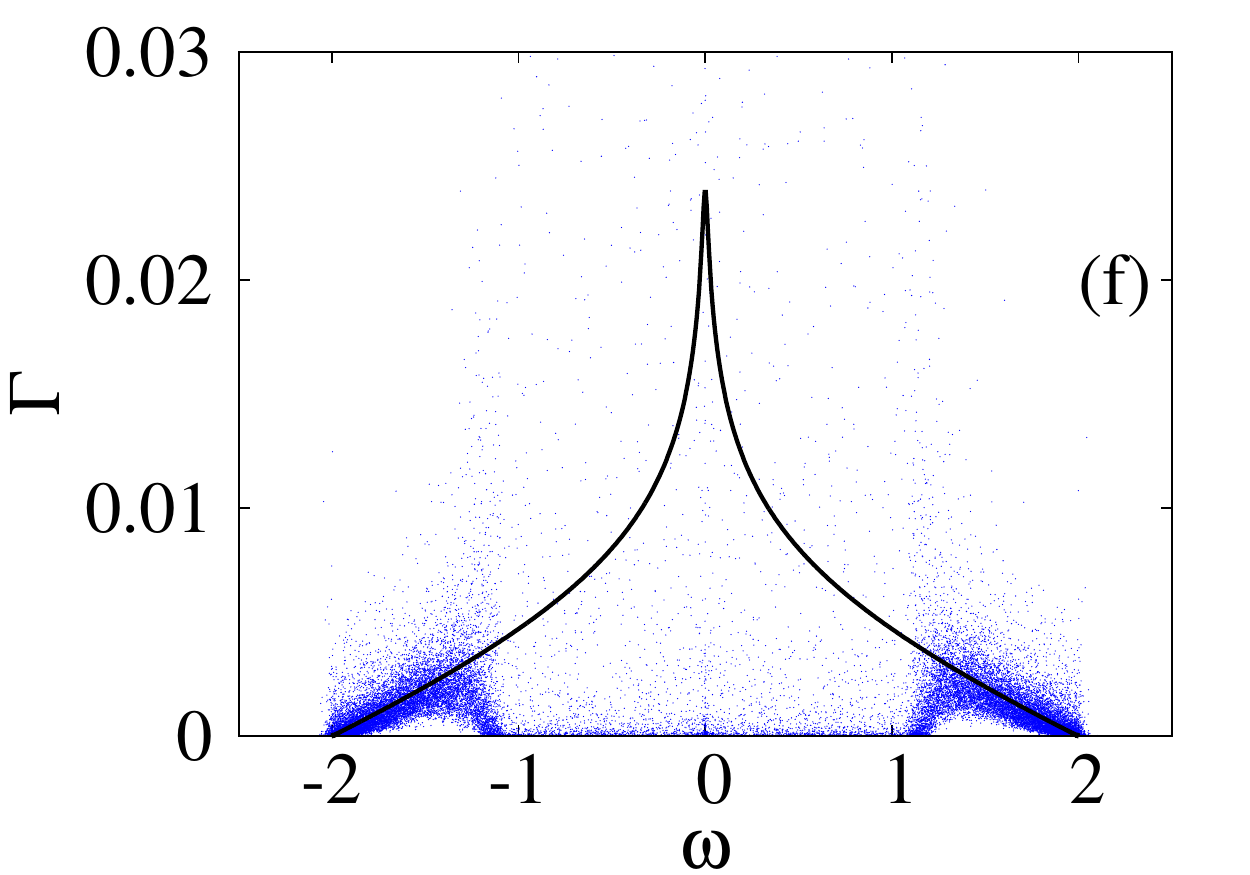}
\caption{\label{Fig8} 
Imaginary vs.~real part of the $S$-matrix poles $\Omega$ [see Eq.~(\ref{poles})] for 1D 
wires with correlated 
disorder defined by the power spectrum (\ref{mu3}).  
Wires of length $N=970$ are coupled to the continuum with strengths $\gamma=0.1$ 
(upper panels) and $\gamma=1$ (lower panels). The disorder strength was set to 
$\sigma^2=0.001$, $L_{\mbox{\scriptsize{loc}}}/N=10$ (left panels); 
$\sigma^2=0.01$, $L_{\mbox{\scriptsize{loc}}}/N \approx 1$ (middle panels); and 
$\sigma^2=0.1$, $L_{\mbox{\scriptsize{loc}}}/N=0.1$ (right panels).
Black curves represent the corresponding non-disorder wires. 
Here, 50 wire realizations were used.}
\end{figure*}

The corresponding distributions of $S$-matrix poles are displayed in Fig.~\ref{Fig8}. In the left 
panels (ballistic regime) the poles whose real values belong to energy intervals where the localization 
length is infinite are practically equal to the corresponding poles of the non-disordered system. We 
term these energy intervals with $\lambda=0$ as windows of transparency, since here the transmission 
of waves is practically perfect. Outside of these windows, where the localization length is finite, a clear 
deviation from the non-disordered case is observed. 

As the ratio $L_{\mbox{\scriptsize{loc}}}/N$ between the localization length and the system size 
decreases, the poles whose real parts belong to extended eigenstates in the windows of transparency 
begin to spread around the corresponding poles of the non-disorder case. The spreading of these 
poles is stronger as $L_{\mbox{\scriptsize{loc}}}/N$ decreases (see middle and right panels in 
Fig.~\ref{Fig8}). With a further decrease of $L_{\mbox{\scriptsize{loc}}}/N$, the eigenstates begin to be 
strongly localized and one can observe an accumulation of poles towards the real axis (see right panels 
of Fig.~\ref{Fig8}).

\subsection{Mobility edges II}

Here we show that the windows of transparency and the regions along the band with strongly localized 
eigenstates can be easily interchanged by the proper choice of long-range correlations. With respect to 
the case discussed in previous Subsection, this can be accomplished by the following power spectrum:
\begin{eqnarray}
     W(E)= 
         \left\{
                \begin{array}{ll}
                    0 \ , & \quad  -E_1<E<E_1 \ , \\
                    \pi/2\mu_1 \ , & \quad  \mbox{otherwise} \ . 
                \end{array}
         \right.
\label{mu4}
\end{eqnarray}
With this type of correlations, the eigenstates whose eigenvalues belong to the energy interval 
$[E_1,E_1]$ are expected to be extended, whereas the eigenvalues located outside of this energy 
window should correspond to localized eigenstates. This prediction is corroborated by the numerical 
data as Fig.~\ref{Fig10}(d) shows. Notice that now the ratio $L_{\mbox{\scriptsize{loc}}}/N$ between the 
localization length and 
the system size evaluated at the band center is infinite. Therefore, strictly speaking, the system is 
in the ballistic regime independently of the system size and disorder intensity. However, as we will 
see below, the disorder intensity still plays an important role in the distribution of poles.  

\begin{figure*}[!htp]
\includegraphics[width=5.5cm,height=4cm]{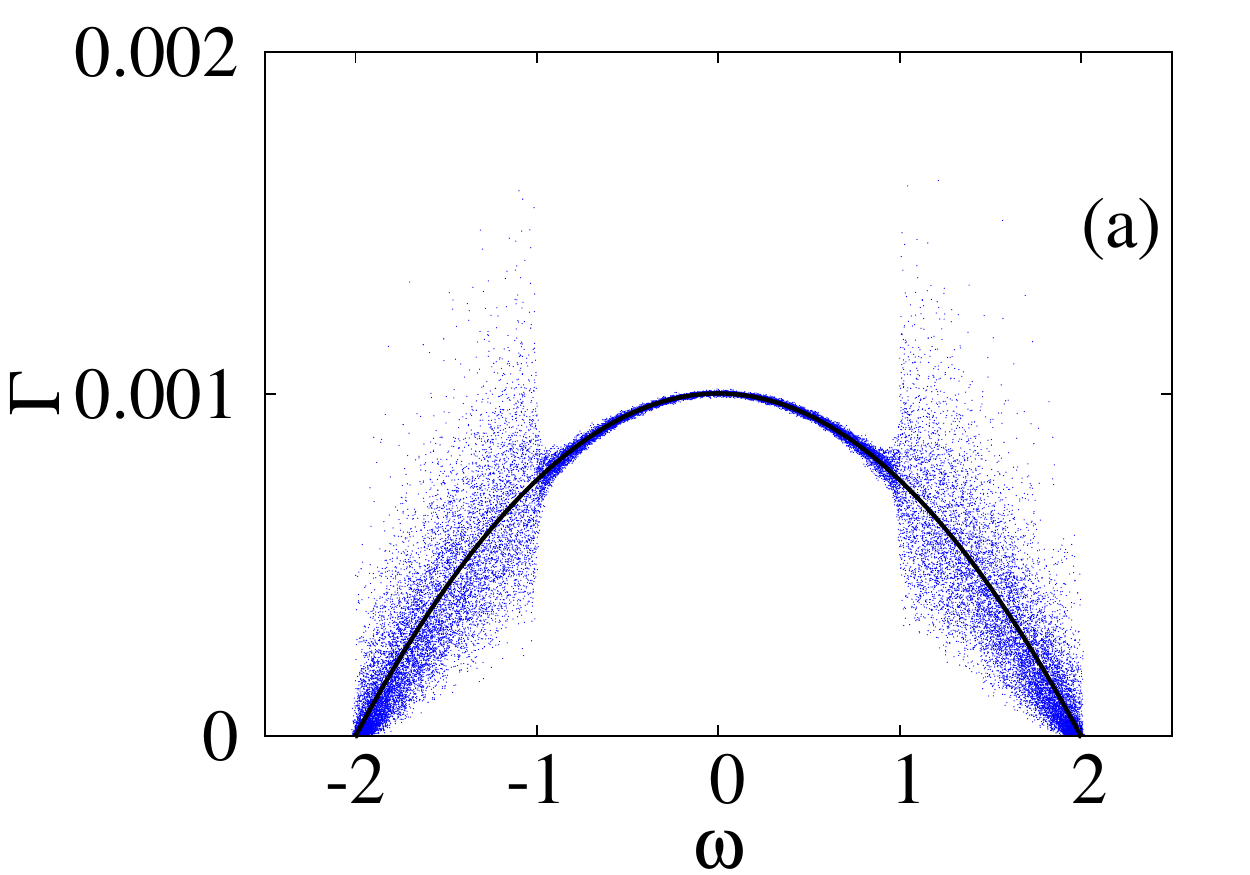}
\includegraphics[width=5.5cm,height=4cm]{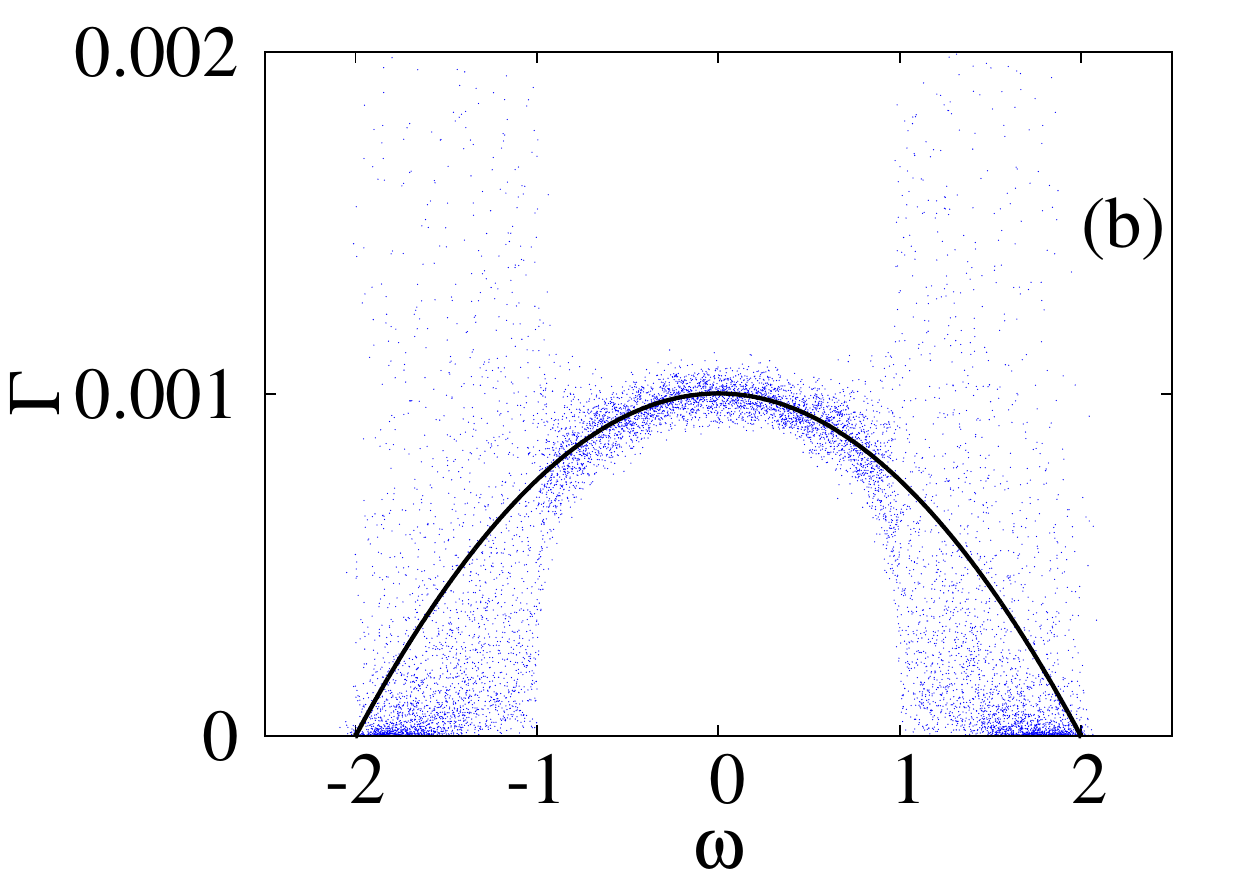}
\includegraphics[width=5.5cm,height=4cm]{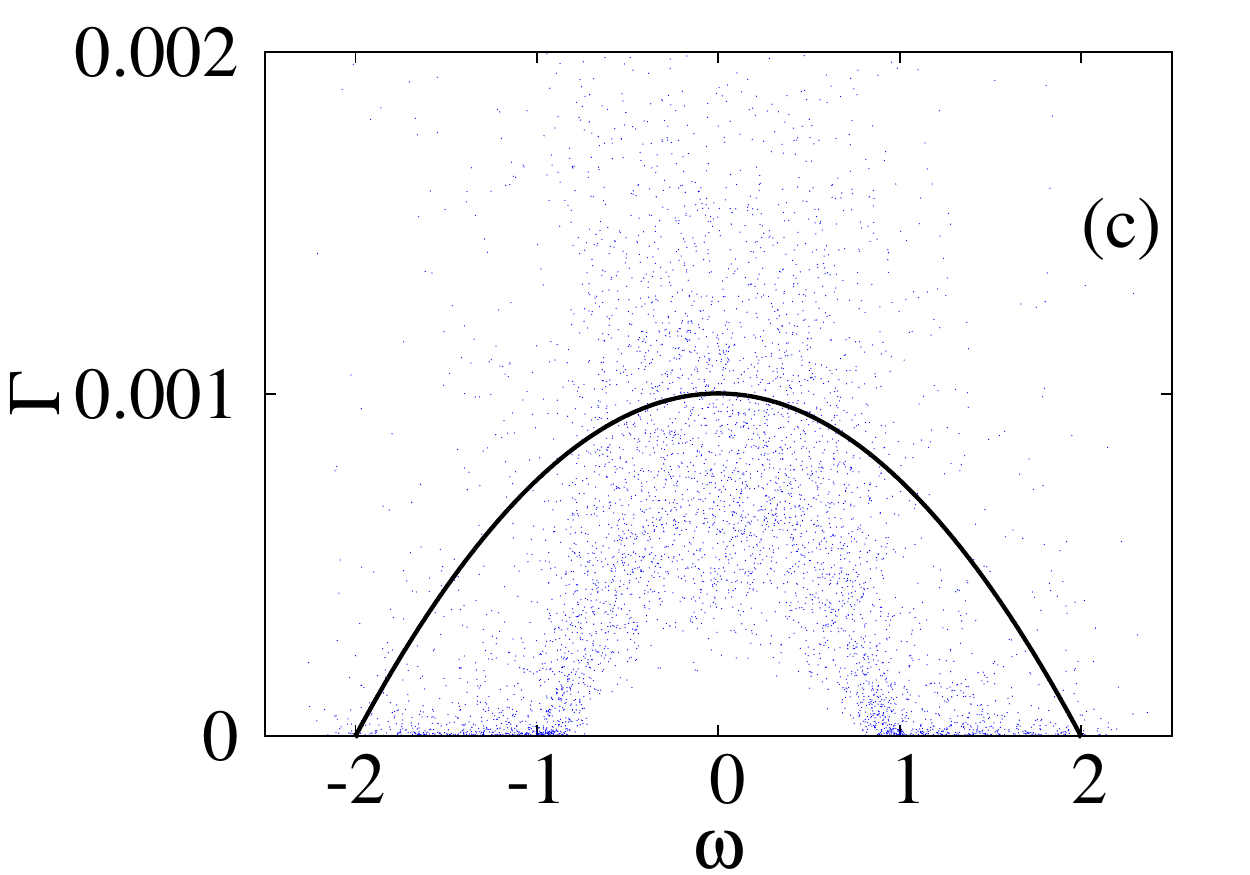}
\includegraphics[width=5.5cm,height=4cm]{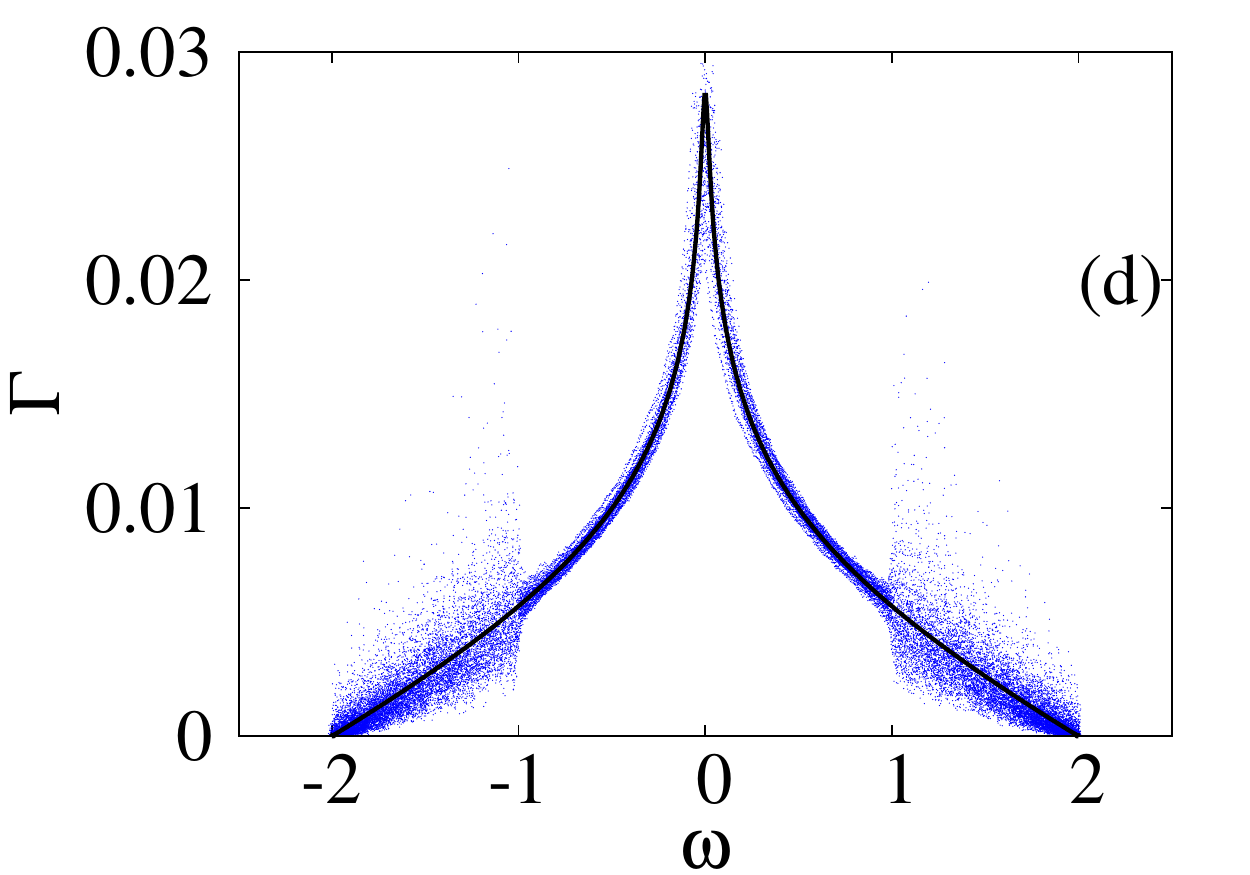}
\includegraphics[width=5.5cm,height=4cm]{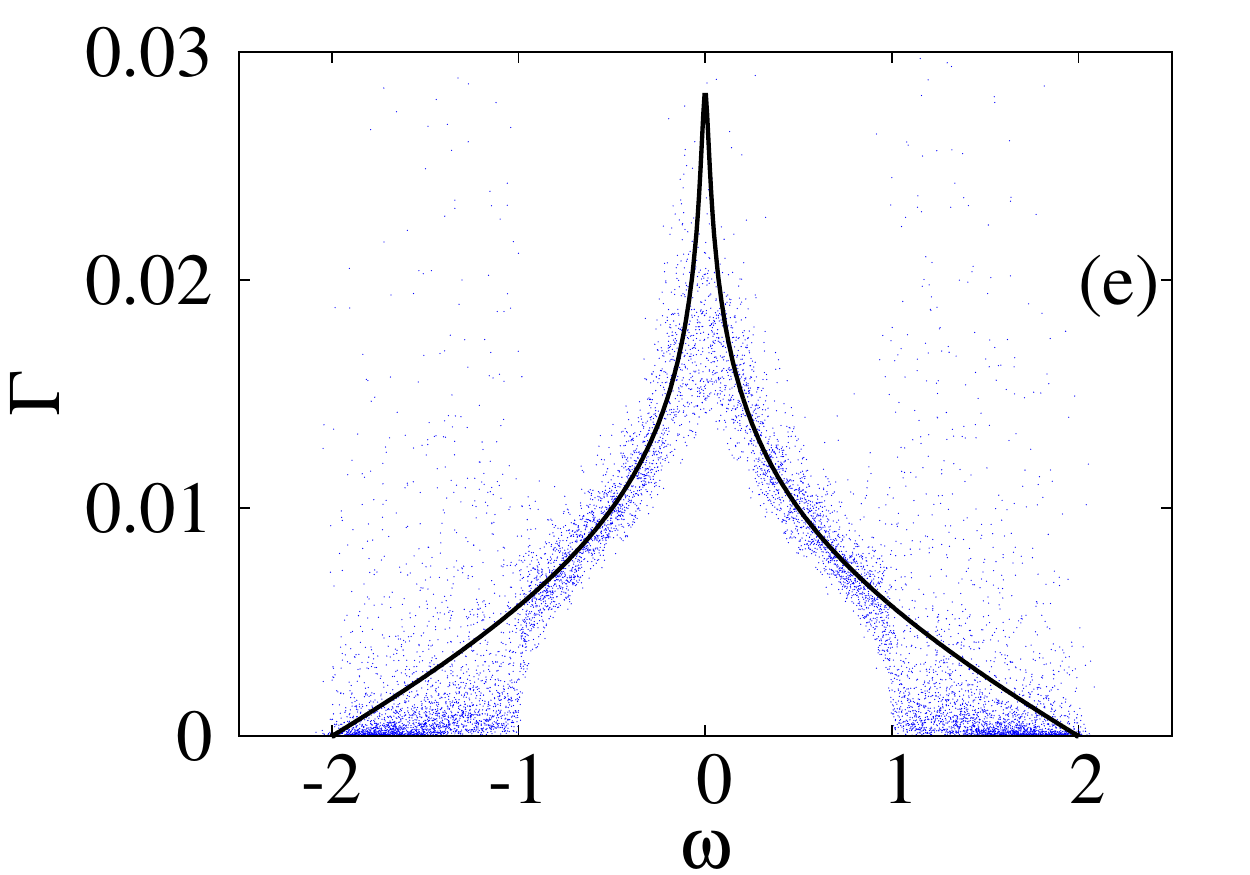}
\includegraphics[width=5.5cm,height=4cm]{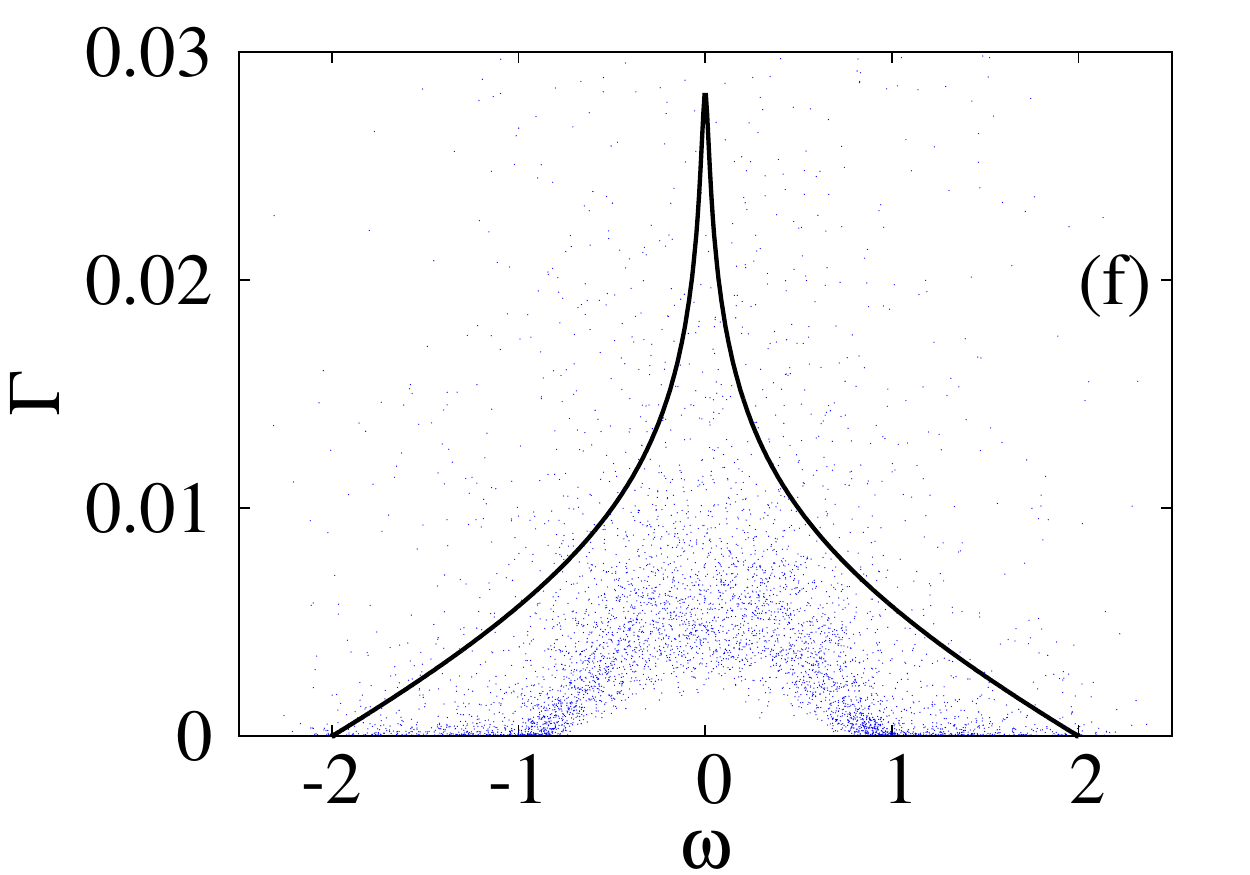}
\caption{\label{Fig9} Imaginary vs.~real part of the $S$-matrix poles $\Omega$ [see Eq.~(\ref{poles})] for 1D 
wires with correlated disorder defined by the power spectrum (\ref{mu4}).  
Wires of length $N=800$ are coupled to the continuum with strengths $\gamma=0.1$ 
(upper panels) and $\gamma=1$ (lower panels). The disorder strength was set to 
$\sigma^2=0.001$, $L_{\mbox{\scriptsize{loc}}}/N=10$ (left panels); 
$\sigma^2=0.01$, $L_{\mbox{\scriptsize{loc}}}/N \approx 1$ (middle panels); and 
$\sigma^2=0.1$, $L_{\mbox{\scriptsize{loc}}}/N=0.1$ (right panels).
Black curves represent the corresponding non-disorder wires. 
Here, 50 wire realizations were used.}
\end{figure*}

The binary correlator that results in the power spectrum of Eq.~(\ref{mu4}) is given by
\begin{eqnarray}
K(m)=\frac{\sin(2\mu_1 m)}{2\mu_1 m} \ ,
\end{eqnarray}
which exhibits a power law decay typical of long-range correlated disorder. The distributions of poles 
of the $S$-matrix are displayed in Fig.~\ref{Fig9}. For very weak disorder, $\sigma^2\lesssim 0.001$ 
(left panels), the poles whose real part belongs to the extended eigenstates are practically distributed 
as for the non-disorder case. As we increase the disorder these poles begin to spread around the 
corresponding poles of the non-disordered wire. In contrast, in the complementary energy windows 
with localized eigenstates most of the poles are located close to the real axis while the rest acquire 
a large imaginary part. Therefore, in these localization windows we have two effects. On the one 
hand the number of poles that are close to the real axis is increased as the disorder becomes 
stronger; on the other hand, the imaginary part of the rest of poles becomes larger (see middle and 
right panels of Fig.~\ref{Fig9}). If the disorder is strong enough, the distribution of poles whose real 
part belongs to the extended eigenstates is no longer similar to that of the poles of the non-disordered 
wire.

%%%%%%%%%%%%%%%%%%%%%%%%%%%%%%%%%%%%%%%%%%%%%%%%%%%%%%%%%%%%%%%%%%%%%%%%%%%%%%%%%%%%%%%%%
%%%%%%%%%%%%%%%%%%%%%%%%%%%%%%%%%%%%%%%%%%%%%%%%%%%%%%%%%%%%%%%%%%%%%%%%%%%%%%%%%%%
%%%%%%%%%%%%%%%%%%%%%%%%%%%%%%%%%%%%%%%%%%%%%%%%%%%%%%%%%%%%%%%%%%%%%%%%%%%%%%%%%%%

\section{Summary}

We have studied 1D open tight-binding disordered wires paying main attention to the distribution of 
poles of the $S$-matrix in dependence on the model parameters. The model essentially depends on 
the system size $N$, the strength $\gamma$ of the coupling to the leads, the square-root-variance 
$\sigma$ of weak diagonal disorder, and on the type of disorder. In the first part of the paper we have 
considered uncorrelated disorder and ask the question of how the pole distribution depends on two 
key parameters. One of these parameters is the ratio $L_{\mbox{\scriptsize{loc}}}/N$ of the localization 
length $L_{\mbox{\scriptsize{loc}}}$ in the corresponding closed system (for $\gamma=0$) to the 
system size $N$. It is known that for perfect coupling ($\gamma = 1$) this ratio determines all transport 
properties of the open system, a fact known as the single parameter scaling in the theory of localization,
see for example Ref.~\cite{M99}. 
According to this parameter, we consider three characteristic situations: extended 
eigenstates with $L_{\mbox{\scriptsize{loc}}} \gg N$  (plane waves slightly modified by disorder), 
extended chaotic eigenstates with $L_{\mbox{\scriptsize{loc}}} \sim N$, and localized eigenstates with 
$L_{\mbox{\scriptsize{loc}}} \ll N$. 

The analysis of the pole distribution in dependence on $L_{\mbox{\scriptsize{loc}}}/N$ and $\gamma$ 
has shown that for the effectively weak disorder ($L_{\mbox{\scriptsize{loc}}} \gg N$) the location of 
poles in the complex plane follows those occurring for a non-disordered potential ($\sigma=0$ and 
$\gamma = 0$). In the other limit case of relatively strong disorder ($L_{\mbox{\scriptsize{loc}}} \ll N$) 
it was found that the pole distribution is very 
different from the previous case. Specifically, the data clearly demonstrates that, when in the closed 
wire the eigenstates are strongly localized, in the open wire the poles are mainly located close to 
the real axis. With the increase of coupling to the leads this effect is enhanced. It is important to 
stress that even for weak disorder, the effect of attraction of the poles to the real axis cannot be 
neglected (see the data in Figs.~\ref{Fig1} and \ref{Fig2}).

Another question is how the mean value of the imaginary parts of the poles depends on the strength 
of the coupling to the leads. Note that the poles of the $S-$matrix give the information about 
the widths of the resonances appearing in the transmission of waves through finite disordered wires. 
These resonances can easily be observed experimentally, at least for the poles with a small 
imaginary part. Our key idea was that the dependence of the mean value of the imaginary parts of 
the poles on the coupling strength is similar to that described by the famous Moldauer-Simonius 
relation (\ref{Eq10}). Although this relation has been derived for random matrix models, one can 
expect that Eq.~(\ref{Eq10}) is also valid for low-dimensional models such as the model studied here. 
The numerical data 
reported in Fig.~\ref{Fig3} clearly support our expectation. Thus our results indicate that the area of 
application of the Moldauer-Simonius relation (\ref{Eq10}) is much broader than it was initially expected.

The most important feature described by the Moladauer-Simonious relation is that at the critical point, 
$\gamma=1$, the mean value of the widths $\Gamma$ diverges. This remarkable fact is well seen in 
Fig.~\ref{Fig3}, also demonstrating an increase of fluctuations of individual $\Gamma$ at the critical point. 
Note that these two effects seem to be independent of the degree of localization, $L_{\mbox{\scriptsize{loc}}}/N$. 
However, the fluctuations themselves are increased with the decrease of this ratio, therefore, 
with the increase of the disorder. It should be stressed that the divergence of the mean of widths at 
$\gamma=1$ can be treated as an indication of a phase transition, for which the fluctuations are of the 
order of the mean values. This effect is known as the superradiance transition well studied in terms of 
full random matrices in place of the real part $H$ in non-Hermitian Hamiltonians \cite{brody81}.

In order to better present the data characterizing the superradiant transition, we have plotted the mean 
values of $\Gamma$ for two poles (for each wire) with the largest values of $\Gamma$. In the region 
$\gamma<1$ there is only one cloud of poles in the complex plane and all the poles have effectively 
small imaginary parts (the region of isolated resonances). 
When the coupling exceeds the critical value $\gamma=1$, two of the $N$ poles 
have very large values of their imaginary parts in comparison with all other poles that move back 
to the real axis with the increase of the coupling. This effect is clearly seen in Fig.~\ref{figu35}, where the 
mean value of the largest $\Gamma $ is plotted. As one can see, for 
$\gamma < 1$ the influence of localization is strong for localized states, however, for $\gamma > 1$ the 
effect of the localization is negligible. 

In the second part of the paper we addressed the question of the influence of correlations, imposed to the 
diagonal disorder, on the pole distribution. We have considered few types of disorder with either short-range 
or long-range correlations. Having in mind the analytical results obtained for all these cases in the weak 
disorder limit, first in Fig.~\ref{Fig10} we plotted the localization length versus energy for the potentials with 
correlations, in comparison with the analytical predictions. Our data demonstrate that the method, used for 
the creation of random disorder resulting in specific dependences of the localization length on energy, 
works perfectly in the presence of coupling to continuum.

With the use of the theory of correlated disorder one can create the 
energy windows where the localization length is much larger that the length $N$ of the wires, together with 
the windows where the eigenstates are strongly localized. In this way, one can speak of effective band edges. 
The data in Fig.~\ref{Fig10} demonstrate an excellent agreement between the numerically found localization 
length and the analytical predictions. However, a quite strong discrepancy occurs for energies 
close to the band center; an effect which is well studied in the literature (see for example the review in 
Ref.~\cite{IKM12}). The origin of this discrepancy is the failure of the standard perturbation theory used 
to derive the analytical expression for the localization length. 

With the reference to Fig.~\ref{Fig10}, where the localization length is plotted versus energy, we have 
analyzed the distribution of poles when correlations are imposed into disorder in the presence of 
coupling to the leads. In Figs.~\ref{Fig6} and \ref{Fig7} the 
pole distribution is shown for correlated disorder without mobility edges in dependence on the disorder 
strength and on the degree of the coupling. The analysis shows that, in general, with the increase of 
disorder the poles begin to be more scattered in comparison to the non-disordered wires. Another 
conclusion is that for both weak and strong disorder the poles tend to be concentrated near the real axis. 
As for the correlated disorder resulting in the mobility edges, see Figs.~\ref{Fig8} and \ref{Fig9}, the most 
important conclusion is that in the presence of coupling to the leads the distribution of the poles of the 
scattering matrix mainly follows that occurring in the absence of disorder, provided the coupling parameter
$\gamma$ is not too large. On the other hand, strong coupling to the continuum essentially modifies the 
distribution of poles, however, mainly in those energy windows where the eigenstates are strongly localized. 

\section{acknowledgments}
We acknowledge G. L. Celardo for pointing out relevant references to us.
F.M.I. acknowledges support from 
VIEP-BUAP (Grant No.~IZF-EXC13-G).
J.A.M.-B. acknowledges support from 
VIEP-BUAP (Grant No.~MEBJ-EXC18-G), 
Fondo Institucional PIFCA (Grant No.~BUAP-CA-169), and 
CONACyT (Grant No.~CB-2013/220624).

\end{document}